\documentclass[10pt,conference]{IEEEtran}
\usepackage{cite}
\usepackage{amsmath,amssymb,amsfonts}
\usepackage{algorithmic}
\usepackage{graphicx}
\usepackage{textcomp}
\usepackage{xcolor}
\usepackage[hyphens]{url}
\usepackage{fancyhdr}

\usepackage{cite}
\usepackage{amsmath,amssymb,amsfonts}
\usepackage{algorithmic}
\usepackage{textcomp}
\usepackage{xcolor}
\usepackage[hyphens]{url}
\usepackage{fancyhdr}
\usepackage[bookmarks=true,breaklinks=true,letterpaper=true,colorlinks,citecolor=blue,linkcolor=blue,urlcolor=blue]{hyperref}
\usepackage{multirow}
\usepackage{booktabs}
\usepackage{tikz}
\usepackage{comment}
\usepackage{array}
\usepackage{subcaption}
\usepackage[flushleft]{threeparttable}
\usepackage{threeparttable}
\usepackage[font=small,labelfont=bf]{caption}
\usepackage{dblfloatfix}
\usepackage{amssymb}
\usepackage{pifont}
\definecolor{ao(english)}{rgb}{0.0, 0.5, 0.0}
\newcommand{\NAME}{CogSys\space}  
\newcommand*\circled[1]{\tikz[baseline=(char.base)]{
            \node[shape=circle,fill,inner sep=0.2pt] (char) {\textcolor{white}{#1}};}}

\newcommand{\hpcayear}{2025}

\newcommand{\hpcasubmissionnumber}{1580}
\title{\fontsize{23.5}{30.0}\selectfont{CogSys: Efficient and Scalable Neurosymbolic Cognition System via Algorithm-Hardware Co-Design \vspace{-25pt}}}

\def\hpcacameraready{} 

\newcommand\hpcaauthors{Zishen Wan$^{\dagger*}$, Hanchen Yang$^{\dagger*}$, Ritik Raj$^{\dagger*}$, Che-Kai Liu$^{\dagger}$, Ananda Samajdar$^{\ddagger}$,\\ Arijit Raychowdhury$^{\dagger}$, Tushar Krishna$^{\dagger}$}
\newcommand\hpcaaffiliation{$^{\dagger}$\textit{Georgia Institute of Technology, Atlanta, GA} \hspace{5pt} $^{\ddagger}$\textit{IBM Research, Yorktown Heights, NY}}
\newcommand\hpcaemail{\{zishenwan, hanchen, ritik.raj, che-kai\}@gatech.edu, ananda.samajdar@ibm.com \\ \{arijit.raychowdhury, tushar\}@ece.gatech.edu}

\author{
  \ifdefined\hpcacameraready
    \IEEEauthorblockN{\hpcaauthors{}}
      \IEEEauthorblockA{
        \hpcaaffiliation{} \\
        \hpcaemail{}
      }
  \else
    \IEEEauthorblockN{\normalsize{HPCA \hpcayear{} Submission
      \textbf{\#\hpcasubmissionnumber{}}} \\
      \IEEEauthorblockA{
        Confidential Draft \\
        Do NOT Distribute!! \vspace{-5pt}
      }
    }
  \fi 
}

\fancypagestyle{camerareadyfirstpage}{%
  \fancyhead{}
  
  \fancyhead[C]{
    \ifdefined\aeopen
    \parbox[][12mm][t]{13.5cm}{\hpcayear{} IEEE International Symposium on High-Performance Computer Architecture (HPCA)}    
    \else
      \ifdefined\aereviewed
      \parbox[][12mm][t]{13.5cm}{\hpcayear{} IEEE International Symposium on High-Performance Computer Architecture (HPCA)}
      \else
      \ifdefined\aereproduced
      \parbox[][12mm][t]{13.5cm}{\hpcayear{} IEEE International Symposium on High-Performance Computer Architecture (HPCA)}
      \else
      \parbox[][0mm][t]{13.5cm}{\hpcayear{} IEEE International Symposium on High-Performance Computer Architecture (HPCA)}
    \fi 
    \fi 
    \fi 
    \ifdefined\aeopen 
      \includegraphics[width=12mm,height=12mm]{ae-badges/open-research-objects.pdf}
    \fi 
    \ifdefined\aereviewed
      \includegraphics[width=12mm,height=12mm]{ae-badges/research-objects-reviewed.pdf}
    \fi 
    \ifdefined\aereproduced
      \includegraphics[width=12mm,height=12mm]{ae-badges/results-reproduced.pdf}
    \fi
  }
  \fancyfoot[C]{}
}
\begin{document}
\maketitle

\ifdefined\hpcacameraready 
  \thispagestyle{camerareadyfirstpage}
  \pagestyle{empty}
\else
  \thispagestyle{plain}
  \pagestyle{plain}
\fi

\newcommand{\hpcaheight}{0mm}
\ifdefined\eaopen
\renewcommand{\hpcaheight}{12mm}
\fi


 \begin{abstract}
Neurosymbolic AI is an emerging compositional paradigm that fuses neural learning with symbolic reasoning to enhance the transparency, interpretability, and trustworthiness of AI. It also exhibits higher data efficiency making it promising for edge deployments. 
Despite the algorithmic promises and demonstrations, unfortunately executing neurosymbolic workloads on current hardware 
(CPU/GPU/TPU) is challenging due to higher memory intensity, greater compute heterogeneity and 
access pattern irregularity, leading to severe hardware underutilization.
%

This work proposes CogSys, a characterization and co-design framework dedicated to neurosymbolic AI system acceleration, aiming to win both reasoning efficiency and scalability. 
\underline{On the algorithm side}, \NAME proposes an \textit{efficient factorization technique} 
to alleviate compute and memory overhead.
\underline{On the} \underline{hardware side}, \NAME proposes a scalable neurosymbolic architecture with \textit{reconfigurable neuro/symbolic processing elements (nsPE)} and \textit{bubble streaming (BS) dataflow} with \textit{spatial-temporal (ST) mapping} for highly parallel and efficient neurosymbolic computation.
\underline{On the system side}, \NAME features an \textit{adaptive workload-aware scheduler (adSCH)} to orchestrate heterogeneous kernels and enhance resource utilization.
Evaluated across cognitive workloads, \NAME 
enables reconfigurable support for neural and symbolic kernels and exhibits $>$75$\times$ speedup over TPU-like systolic array with only $<$5\% area overhead, as benchmarked under the TSMC 28nm technology node.
\NAME achieves 4$\times$-96$\times$ speedup compared to desktop and edge GPUs.
For the first time, \NAME enables real-time \textit{abduction reasoning} towards human fluid intelligence, requiring only 0.3~s per reasoning task
with 4~mm$^2$ area and 1.48~W power consumption. 



\end{abstract}

\section{Introduction}
\label{sec:intro}


The massive success of Large Language Models (LLMs) combined with concerns about interpretability and safety have led to an emerging paradigm of ``compositional AI" systems - especially for safety-critical domains such as robotics and healthcare. 
The goal of such systems is to combine black-box neural networks with reasoning/rule-based AI methods~\cite{wan2024towards,murule,zhao2024retrieval,kang2024r,kwon2024neuro,kalyanpur2024llm,xiong2024converging,ibrahim2024special}.
This approach mirrors human cognitive processes, which can be grouped as lower-level sensory processing (system 1 ``thinking fast'') and higher-level cognitive functions like reasoning and deduction (system 2 ``thinking slow'')~\cite{daniel2017thinking,booch2021thinking}. The former can be modeled with neural networks, and the latter with symbolic frameworks.\footnotetext{$^{*}$Equal Contributions.}

\begin{figure}[t!]
\centering
\includegraphics[width=\columnwidth]{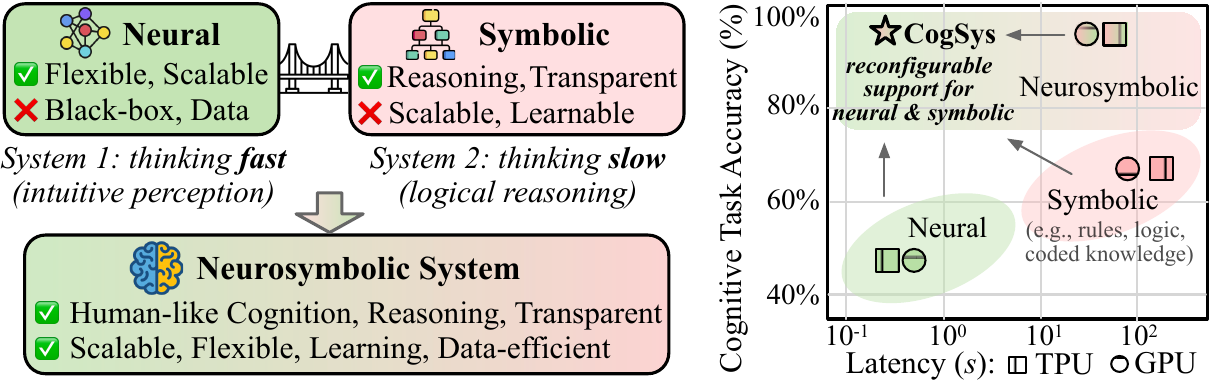}
\vspace{-0.2in}
        \caption{\textbf{Neurosymbolic AI} is an emerging \textit{compositional} system that integrates neural and symbolic modules, enabling superior cognitive intelligence compared to NNs. However, it suffers from inefficient TPU/GPU execution. \textbf{CogSys} is a reconfigurable neural/symbolic engine excelling in both reasoning efficiency and cognitive capability.}
        \label{fig:nesy_intro}
        \vspace{-12pt}
\end{figure}

One promising example of compositional AI system is \textbf{neurosymbolic AI} that synergistically integrates neural network learning with \textit{symbolic reasoning}. The neural networks are adept at identifying patterns and handling perceptual tasks, but lack transparency and logical inference capabilities. Symbolic modules (e.g., coded knowledge, rules), in contrast, excel in reasoning and interpretability but struggle with adaptability and learning from raw data. Neurosymbolic AI bridges this gap by composing strengths of both paradigms (Fig.~\ref{fig:nesy_intro}).
%

Neurosymbolic AI has
demonstrated superior capabilities in human-like reasoning and logical thinking across various domains, such as natural language processing, robotics, healthcare, etc~\cite{garcez2023neurosymbolic,wan2024towards,mao2019neuro,han2019visual,mei2022falcon,yi2020clevrer,zhang2021abstract,shah2022knowledge,hsu2023ns3d,hersche2023neuro}. 
For example, IBM's neuro-vector-symbolic system~\cite{hersche2023neuro} achieves 98.8\% accuracy on spatial-temporal reasoning tasks~\cite{raven}, greatly surpassing human (84.4\%), ResNet (53.4\%) and GPT-4 (89.0\%); Google DeepMind's AlphaGeometry~\cite{trinh2024solving}, another neurosymbolic system, solves geometry problems at a level of human Olympiad gold medalists, while GPT-4 completely fails. 
Recently, there also has been a plethora of workshops focusing on neurosymbolic AI~\cite{neus24,nesy24,aaai24-workshop,aaai23-tutorial,neurips24-workshop,ibm23-workshop,nesy23-summer,ijcai24-workshop}.

\begin{figure*}[t!]
\centering
\includegraphics[width=2.05\columnwidth]{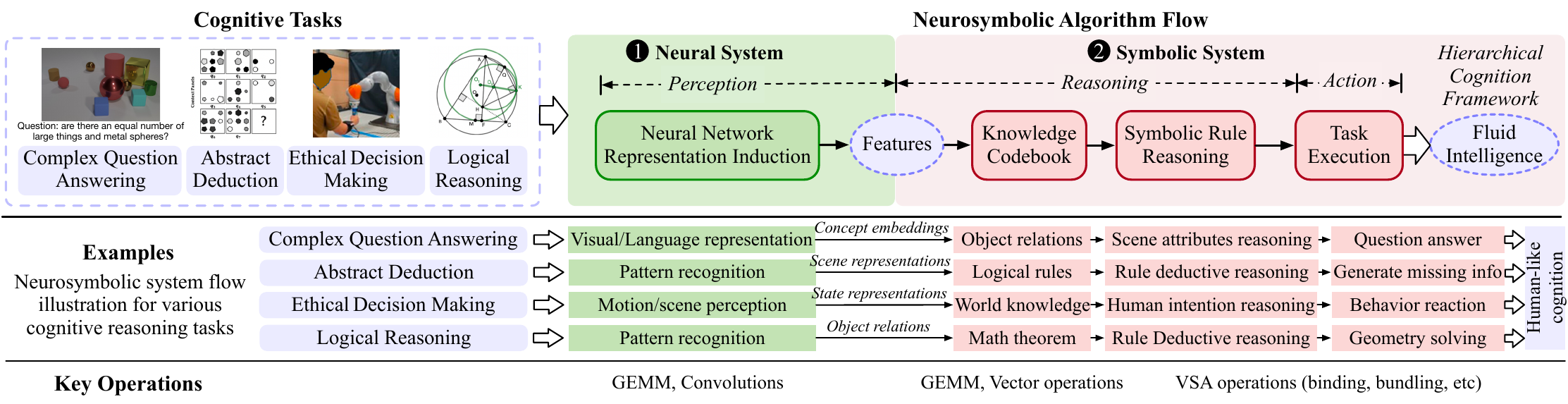}
\vspace{-0.2in}
        \caption{\textbf{Neurosymbolic algorithm flow.} Neural systems handle perception by processing raw data and extracting features, which are then utilized by symbolic reasoning systems to apply logical rules and knowledge. This compositionality enables the execution of complex cognitive tasks such as abstract deduction, ethical decision-making, and fluid intelligence.}
    \label{fig:algo_framework}
        \vspace{-10pt}
\end{figure*}

Despite impressive cognitive capabilities of neurosymbolic AI - demonstrated by past work over distributed GPU clusters, 
recent study~\cite{wan2024towards_tcasai} identifies that enabling \textit{real-time and efficient} neurosymbolic AI over edge devices, which is highly desirable for numerous reasoning and human-AI applications, is a challenging open problem.
For example, a neuro-vector-symbolic system takes $>$3~mins even on TPU and desktop GPU for a single task~\cite{hersche2023neuro}. This inefficiency threatens to hurt neurosymbolic AI deployment in the long run.

To understand this further, we systematically profile and analyze the runtime and memory behavior of various neurosymbolic workloads on multiple devices
and identify the following system-level challenges. 
\circled{1} \textbf{Large Memory Footprint.} Neurosymbolic AI systems typically rely on vector-symbolic architecture (VSA) that utilizes vector operations to represent symbolic knowledge.
The system generates an intermediate codebook that captures vast object combinations 
for higher reasoning capability (typically in the order of tens to hundreds MB)
making it impractical to be cached on-chip in edge accelerators.
\circled{2} \textbf{Compute Heterogeneity.} Rather than general matrix multiplications (GEMMs) and convolution operations that current neural hardware largely focuses on accelerating, neurosymbolic workloads typically consist of numerous holographic vector operations (e.g., circular convolution) that run inefficiently on GPU and neural engines like TPUs due to low data reuse, low compute array utilization and low parallelism.
\circled{3} \textbf{Sequential Processing.} Typically, the symbolic-reasoning computation depends on the output of the neuro-perceptual modules, increasing the critical path during cognitive inference and underutilizing parts of the accelerator.

To address the aforementioned challenges, we develop an algorithm-hardware co-design framework, dubbed CogSys, which to the best of our knowledge is the \emph{\textbf{first}} to achieve real-time efficiency and scalability of cognitive neurosymbolic systems, making it more deployable and facilitate neurosymbolic AI development.
\textbf{On the algorithm level,} \NAME proposes an \textit{efficient factorization scheme} to reduce memory footprint. This technique completely replaces the large-size symbolic knowledge codebook, by quickly
factorizing vectors in an interactive manner when decomposing symbolic representations. 
%
%
\textbf{On the hardware level,} \NAME proposes a scalable architecture with \textit{reconfigurable neuro/symbolic processing elements (nsPE)} and \textit{bubble streaming (BS) dataflow} with \textit{spatial-temporal (ST) mapping} for highly parallel and energy-efficient neurosymboic computation. The design is flexible to support heterogeneous neural and circular convolution symbolic kernels across vector dimensions and reduce runtime.
\textbf{On the system level,} \NAME also features an \textit{adaptive workload-aware scheduling (adSCH) scheme} with \textit{multi-level parallelism} to orchestra neural and symbolic kernels with improved hardware resource utilization and enables design scalability for evolving neurosymbolic AI. \emph{Notably, with only $<$5\% area overhead over TPU-like systolic arrays, \NAME enables reconfigurable support for neural and symbolic kernels and demonstrates $>$75$\times$ system speedup.}


This paper, therefore, makes the following contributions:


\textbullet~We perform comprehensive runtime and memory analysis of various neurosymbolic workloads across devices, and identify the primary cause of the inefficiency and optimization opportunities, which can also shed light on future neurosymbolic systems acceleration and innovations (Sec.~\ref{sec:profile}).

\textbullet~We propose an algorithm-hardware co-design framework, dubbed CogSys, which is the first to enable real-time, efficient, and scalable VSA-based neurosymbolic systems, making it more deployable and facilitate neurosymbolic AI development.

\textbullet~CogSys innovates across the algorithm-level efficient symbolic factorization strategy (Sec.~\ref{sec:algo_opt}), hardware-level reconfigurable neuro/symbolic architecture and dataflow (Sec.~\ref{sec:accelerator}), and system-level scheduler (Sec.~\ref{sec:sched}) to reduce the memory footprint while improving hardware utilization and overall neurosymbolic processing efficiency.




\textbullet~Evaluated across cognitive tasks, \NAME enables reconfigurable support for neural and symbolic operations, achieving 75.9$\times$ speedup with only a 4.8\% area overhead compared to TPU-like systolic arrays, and demonstrates 4$\times$-95$\times$ speedup compared to GPUs.
\NAME enables efficient neurosymbolic AI with 4mm$^2$ area and 1.48W power consumption (Sec.~\ref{sec:eval}). 
\section{Neurosymbolic AI Background and Workload}
\label{sec:background}

This section presents the preliminaries of neurosymbolic AI with its algorithm flow and key operations, then describes four representative neurosymbolic workloads for our analysis.



\vspace{-2pt}
\subsection{Challenges with Neural Networks}
Neural methods are highly effective in extracting complex features from vision and language tasks, and excel in flexibility, scalability, and handling inconsistency~\cite{radford2021learning,zhou2022learning,xiao2023unsupervised,wang2023image,wang2024computation}.
However, neural methods often suffer from limitations such as hallucinations and lack of interpretability, and typically operate as black-box where their decision-making processes are not easily understandable by humans. This undermines the model output trustworthiness in cognitive and safety-critical applications~\cite{wan2021analyzing,ji2023survey,hassija2024interpreting}.

\vspace{-2pt}
\subsection{Neurosymbolic AI Algorithm Flow}
\label{subsec:nsai}
Neurosymbolic AI synergistically integrates the learning capability of neural networks with the reasoning capability of symbolic AI, offering advantages in data-efficient learning with transparent and logical decision-making compared to DNNs. Neurosymbolic AI
leads to superior performance in a wide range of applications, such as complex question answering~\cite{mao2019neuro,mei2022falcon}, abstract deduction~\cite{zhang2021abstract,hersche2023neuro}, decision making~\cite{sheth2024neurosymbolic,nunez2024review}, logical reasoning~\cite{trinh2024solving,romera2024mathematical} tasks,
serving as a promising paradigm to achieve human-like fluid intelligence. 

Fig.~\ref{fig:algo_framework} extracts a unified neurosymbolic pipeline and illustrates how they interact to perform complex cognitive tasks:

\circled{1} \textbf{Neural system.}
The process begins with the neural module handling perception tasks by interpreting sensory data and generating meaningful scene and object representations which are essential for further reasoning processes. The neural module itself may suffer from superposition catastrophe, preventing it from extracting object constituent attributes~\cite{hersche2023neuro}. 


\circled{2} \textbf{Symbolic system.}
The extracted features are fed into the symbolic system for reasoning tasks. This step enhances explainability and reduces dependence on extensive training data by incorporating established models of the physical world (e.g., underlying rules, coded knowledge). Throughout this process, a knowledge codebook is maintained, which integrates learned knowledge from the neural network with symbolic rules, ensuring that the system can both learn from new data and apply logical reasoning based on existing knowledge. The outcomes of symbolic reasoning process are then used to make decisions, generates responses, or controls actions.

This neurosymbolic flow is one way to model human hierarchical reasoning procedures.
Resembling the sense-reason-act cognitive cycle can be computationally modeled through a multi-layer framework~\cite{olascoaga2021brain,ibrahim2024efficient}, where \textit{perception} layer fuses sensory inputs and maps them to high-level observations, \textit{reasoning} layer conducts deliberate and conscious thinking by applying symbolic rules and knowledge, \textit{action} layer facilitates trustworthy and reliable execution. This compositional approach allows agents to tackle complex challenges that require both data-driven learning and logical reasoning.

\subsection{VSA-Based Symbolic Operations}
\label{subsec:vsa_background}

\begin{figure}[t!]
\centering
\includegraphics[width=\columnwidth]{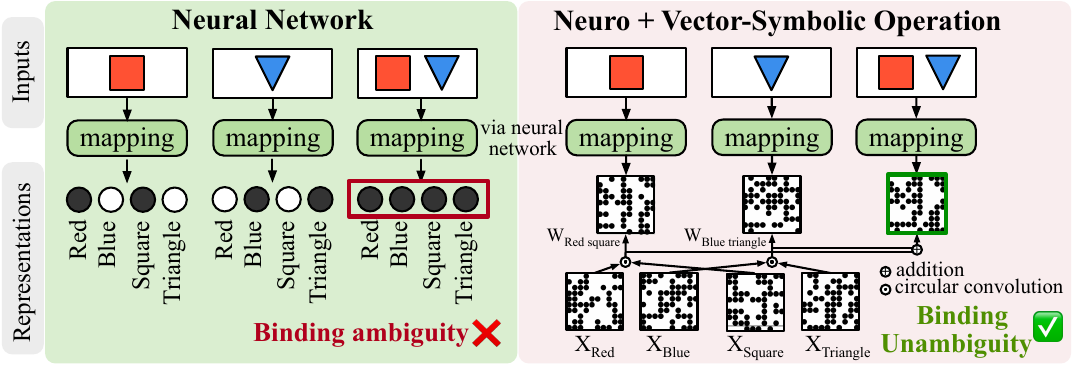}
\vspace{-0.16in}
        \caption{\textbf{Illustration of VSA functionality.} Neural network suffers from binding ambiguity issues, whereas VSA constructs vector representations with circular convolution operations for reasoning process.}
        \label{fig:vsa}
        \vspace{-15pt}
\end{figure}

\textbf{Vector-symbolic architecture (VSA).}
Within the compositional neurosymbolic AI flow, exploiting VSA with neural dynamics has become the powerful approach~\cite{kleyko2022vector,hersche2023neuro,hersche2024probabilistic,furlong2023modelling,zhang2021abstract,menet2024mimonets}.   
Specifically, VSA provides a means to represent symbolic information in a low or high-dimensional vector space. By encoding symbolic structures as vectors with dimensionality-preserving algebraic operations, VSAs enable the combination of symbolic reasoning with neural networks, thereby facilitating cognitive tasks such as learning, memory, and reasoning in a unified system. Fig.~\ref{fig:vsa} illustrates a simple example of the binding ambiguity of neural networks and the functionality of VSA structures.

\textbf{Circular convolution.}
A key VSA operation is the blockwise \textit{circular convolution}, serving as a universal operation for vectors representing different symbols. Circular convolution combines two vectors in a way that preserves the information from both, making it suitable for representing composite symbols. Mathematically, the circular convolution of two vectors \( \mathbf{A} \) and \( \mathbf{B} \) (each of dimension \( N \)) generates vector \( \mathbf{C} \) as $C[n] = \sum_{k=0}^{N-1} A[k] \cdot B[(n-k) \mod N]$ where each element of \( \mathbf{C} \) is obtained by multiplying the elements of \( \mathbf{A} \) with the circularly shifted elements of \( \mathbf{B} \), and then summing up. This process is repeated for each element \( n \) (\( 0 \) to \( N-1 \)). Circular convolution has commutativity and associativity properties, making it particularly effective in hierarchical reasoning tasks where manipulating structured information is critical.

\textbf{Symbolic knowledge codebook.} Symbolic knowledge is typically represented as a set of codebooks for the attributes of interest (Fig.~\ref{fig:vsa}). 
To describe an object with various attributes, a product vector can be computed by binding knowledge codebooks via circular convolution. Due to the properties of multiplicative binding, the co-activated VSA representations result in minimal interference, allowing each object to be accurately recovered. The query vector generated from neural networks will be compared with all codebook vectors to derive attributes for further reasoning. The codebook is typically in the order of tens to hundreds of MB, making it impractical to be cached on-chip in edge accelerators for complex tasks.




\begin{table*}[t!]
\vspace{-0.1in}
\centering
\caption{\textbf{Neurosymbolic models.} 
Selected neurosymbolic AI workloads for analysis, representing a diverse of application scenarios.}
\Huge
\renewcommand*{\arraystretch}{1.15}
\resizebox{\linewidth}{!}{%
\centering
\setlength\tabcolsep{5pt}
\begin{tabular}{cc|c|c|c|c}
\hline
\multicolumn{2}{c|}{\textbf{\begin{tabular}[c]{@{}c@{}}Representative Neuro- \\ Symbolic AI Workloads\end{tabular}}}      
& \textbf{\begin{tabular}[c]{@{}c@{}}Neuro-Vector-Symbolic \\ Architecture~\cite{hersche2023neuro}\end{tabular}}                                                              & \textbf{\begin{tabular}[c]{@{}c@{}}Multiple-Input-Multiple-Output \\ Neural Networks~\cite{menet2024mimonets}\end{tabular}}                                                         & \textbf{\begin{tabular}[c]{@{}c@{}}Probabilistic Abduction via Learning \\Rules in Vector-symbolic Architecture~\cite{hersche2024probabilistic}\end{tabular}}                                                                                           & \textbf{\begin{tabular}[c]{@{}c@{}}Probabilistic Abduction\\ and Execution Learner~\cite{zhang2021abstract}\end{tabular}}                                                                                                                                                          \\ \hline
\multicolumn{2}{c|}{\textbf{Abbreviation}}                                                                                                               & NVSA                                                                                                                                                                                                                                                                                 & MIMONet                                                                                                                                                                    & LVRF                                                                                                                                                                                       & PrAE                                                                                                                                                                                                                                               \\ \hline
\multicolumn{2}{c|}{\textbf{Learning Approach}}                                       & \begin{tabular}[c]{@{}c@{}}Supervised/Unsupervised\end{tabular}                                                                                                                                               & Supervised                                                                                                                                                 & Supervised                                                                                                                                                                                  & \begin{tabular}[c]{@{}c@{}}Supervised/Unsupervised\end{tabular}                                                                                                                                                                                \\ \hline
\multicolumn{1}{c|}{\multirow{2}{*}{\textbf{\begin{tabular}[c]{@{}c@{}}Compute \\ Pattern\end{tabular}}}} & \textbf{Neuro}                                                                                                                                                                                                                                                                                                                 & CNN                                                                                                                                                                                                                     & CNN/Transformer                                                                                                                                                                      & CNN                                                                                                                                                                                       & CNN                                                                                                                                                                                                                                               \\ \cline{2-6} 
\multicolumn{1}{c|}{}                                                                                         & \textbf{Symbolic}                                                                                                                                                                                                                                                                                                           &     VSA binding/unbinding (Circular Conv)
&           VSA binding (Circular Conv) 
&  VSA binding/unbinding (Circular Conv)
& Probabilistic abduction                                                                                 \\ \hline
\multicolumn{1}{c|}{\multirow{3}{*}{\textbf{\begin{tabular}[c]{@{}c@{}}Application\\ Scenario\end{tabular}}}}  & \textbf{Use Case}                                                                                       & \begin{tabular}[c]{@{}c@{}}Spatial-temporal reasoning, Fluid \\intelligence, Abstract reasoning\end{tabular}                                                                                                                             & \begin{tabular}[c]{@{}c@{}}Multi-input simultaneously processing \\ with single CNN/Transformer\end{tabular}   & \begin{tabular}[c]{@{}c@{}}Probabilistic reasoning, Analogy reasoning,\\ Out-of-distribution (OOD) data processing\end{tabular}                                                           & \begin{tabular}[c]{@{}c@{}}Spatial-temporal reasoning, Fluid \\intelligence, Abstract reasoning\end{tabular}                                                                                                                                                           \\ \cline{2-6} 
\multicolumn{1}{c|}{}                                                                                         & \textbf{\begin{tabular}[c]{@{}c@{}}Advantage vs. \\ Neural Model\end{tabular}}                                                                                                                 & \begin{tabular}[c]{@{}c@{}} Higher joint representation efficiency,\\ Better reasoning capability, Transparency\end{tabular}                                                                   & \begin{tabular}[c]{@{}c@{}}Higher throughput, Lower latency, \\ Compositional compute, Transparency\end{tabular}                        & \begin{tabular}[c]{@{}c@{}}Stronger OOD handling capability, One-pass\\ learning, Higher flexibility, Transparency\end{tabular} & \begin{tabular}[c]{@{}c@{}}Higher generalization, Transparency, \\Interpretability, Robustness\end{tabular} \\ \cline{2-6} 
\hline
\end{tabular}
}
\label{tab:selected_model}
\vspace{-5pt}
\end{table*}

\subsection{Representative Neurosymbolic AI Models}
\label{subsec:nsai_model}


Following the flow in Fig.\ref{fig:algo_framework}, we analyze four VSA-based neurosymbolic workloads in detail: NVSA\cite{hersche2023neuro} for spatial-temporal reasoning, MIMONet for multi-input simultaneous processing~\cite{menet2024mimonets}, LVRF for probabilistic abduction~\cite{hersche2024probabilistic}, and PrAE for abstract reasoning~\cite{zhang2021abstract}. These workloads achieve state-of-the-art performance and unlock advanced reasoning capabilities. Our goal is to understand their system and architectural challenges to enable scalable neurosymbolic deployment, where latency and efficiency are critical factors.

\begin{figure*}[t!]
\centering
\vspace{-0.03in}
\includegraphics[width=2.05\columnwidth]{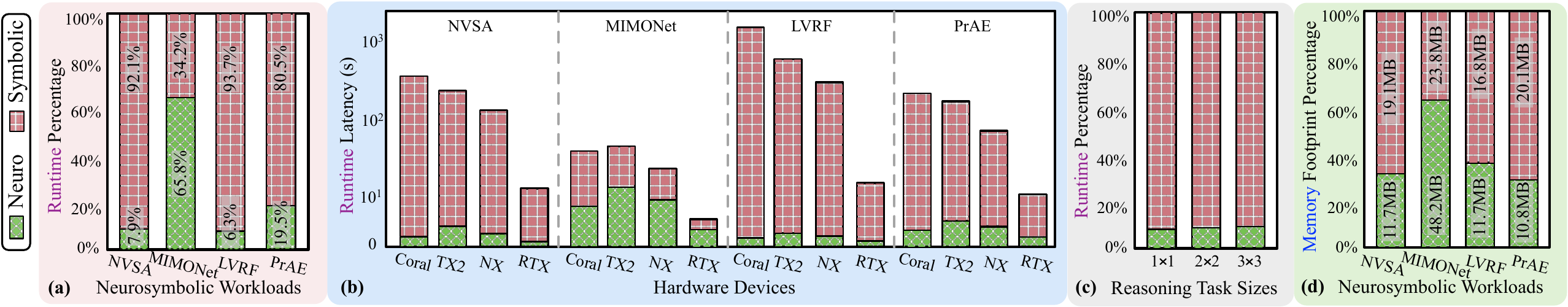}
\vspace{-0.16in}
        \caption{\textbf{End-to-end neurosymbolic runtime, memory, and roofline characterization.} \textbf{(a)} \textnormal{Benchmark neurosymbolic models on CPU+GPU system, showing symbolic may serve as system bottleneck. \textbf{(b)} Benchmark neurosymbolic models on Coral TPU, TX2, NX, and 2080Ti GPU, showing that real-time performance cannot be satisfied. \textbf{(c)} Benchmark models on various task sizes, indicating the potential scalability problem.} \textbf{(d)} Benchmark memory footprint of neurosymbolic models, showing large memory overhead of symbolic knowledge codebook.}
        \label{fig:profiling_latency_e2e}
        \vspace{-12pt}
\end{figure*}

Tab.~\ref{tab:selected_model} lists the details of selected representative workloads:

\subsubsection{Neuro-Vector-Symbolic Architecture (NVSA)}
\label{subsec:nvsa-model}
NVSA is a neurosymbolic system advancing spatial-temporal abduction reasoning~\cite{hersche2023neuro}. Its \emph{neural} module handles visual perception, while the \emph{symbolic} module uses VSA-based operations for probabilistic inference, symbolic rule reasoning, and execution. NVSA bypasses the superposition catastrophe~\cite{rachkovskij2001binding} and surpasses neural-only methods, achieving human-level performance on key fluid intelligence reasoning tests~\cite{raven}.



\subsubsection{Multiple-Input-Multiple-Output Networks (MIMONet)}
MIMONet is a neurosymbolic model designed to handle multiple inputs and reduce computational cost per input~\cite{menet2024mimonets}. Its \textit{neural} modules use CNN/Transformer architectures, while its \textit{symbolic} modules employ VSA binding/unbinding for encoding/decoding, enabling computation in superposition. MIMONet achieves 2-4$\times$ speedup with higher accuracy on LRA benchmarks compared to neural-only methods~\cite{tay2020long}.




\subsubsection{Probabilistic Abduction via Learning Rules in Vector-symbolic Architectures (LVRF)}
LVRF is a neurosymbolic architecture for visual reasoning and handling out-of-distribution data~\cite{hersche2024probabilistic}. Its \textit{neural} modules handle visual perception, while \textit{symbolic} modules use VSA for probabilistic abduction reasoning. LVRF outperforms neural-only methods in unseen reasoning tasks~\cite{iraven}, offering greater flexibility and interpretability.



\subsubsection{Probabilistic Abduction and Execution (PrAE) Learner}
PrAE is a neurosymbolic learner for spatial-temporal cognitive reasoning~\cite{zhang2021abstract}. Its \textit{neural} modules handle visual perception and produce scene representations, while the \textit{symbolic} modules conduct probabilistic reasoning and abduct rules. PrAE offers human-level generalizability, transparency, and interpretability, which classic neural networks struggle to achieve.

\section{Neurosymbolic AI System Profiling}
\label{sec:profile}

Building upon prior profiling study~\cite{wan2024towards}, this section characterizes the system behavior of various vector-symbolic-based neurosymbolic models (Sec.~\ref{subsec:profiling_setup}-\ref{subsec:profiling_symbolic}), and provides workload insights for computer architects (Sec.~\ref{subsec:profile_summary}, \ref{subsec:opportunity}).

\vspace{-2pt}
\subsection{Profiling Setup}
\label{subsec:profiling_setup}
To understand the real-device efficiency of neurosymbolic AI workload, we profile four representative models as elaborated in Sec.~\ref{subsec:nsai_model}, in terms of runtime, memory, and compute operators, for solving cognitive reasoning problems on four 
devices, including Coral edge TPU (4~W), Jetson TX2 (15~W), Xavier NX (20~W), and RTX 2080Ti (250~W), respectively.

\vspace{-2pt}
\subsection{Compute Latency Analysis}
\label{subsec:profiling_latency}

\textbf{End-to-end latency breakdown.}
Fig.~\ref{fig:profiling_latency_e2e}\textcolor{blue}{a} and Fig.~\ref{fig:profiling_latency_e2e}\textcolor{blue}{b} illustrate the end-to-end latency breakdown of four neurosymbolic workloads. We can observe that 
(1) \textit{The real-time performance cannot be satisfied} on all four devices. Even if more computing resources are available to reduce NN runtime, the significant overhead of symbolic reasoning still prohibits real-time execution.
(2) \textit{Symbolic operations consistently dominate runtime.} For example, the symbolic modules count for 87\% of total NVSA inference time while its floating-point operations (FLOPS) count for only 19\% of total NVSA FLOPS, indicating that the symbolic operations may not be well executed by GPU/TPU.
(3) \textit{Symbolic reasoning computation lies on the critical path} due to its dependence on the neuro workloads.

\begin{figure}[t!]
\vspace{-2pt}
  \centering
  \begin{minipage}[b]{0.47\columnwidth}
    \centering
    \includegraphics[width=.85\textwidth]{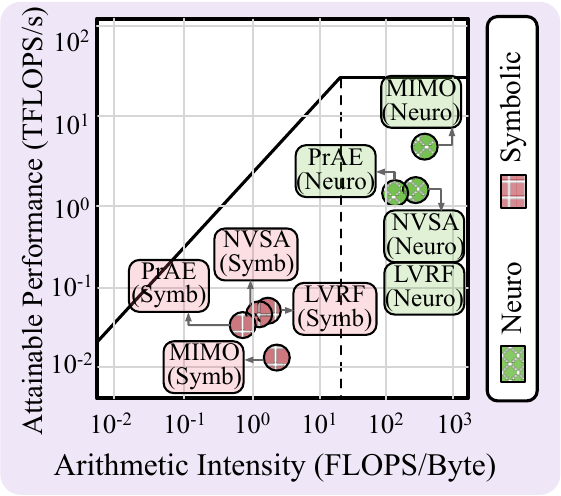}
    \caption{\textbf{Roofline analysis.} End-to-end neurosymbolic roofline characterization on RTX 2080Ti GPU, indicating that typically neuro is compute-bounded and symbolic is memory-bounded.}
    \label{fig:roofline}
  \end{minipage}
  \hfill
  \begin{minipage}[b]{0.49\columnwidth}
    \centering
    \includegraphics[width=.84\textwidth]{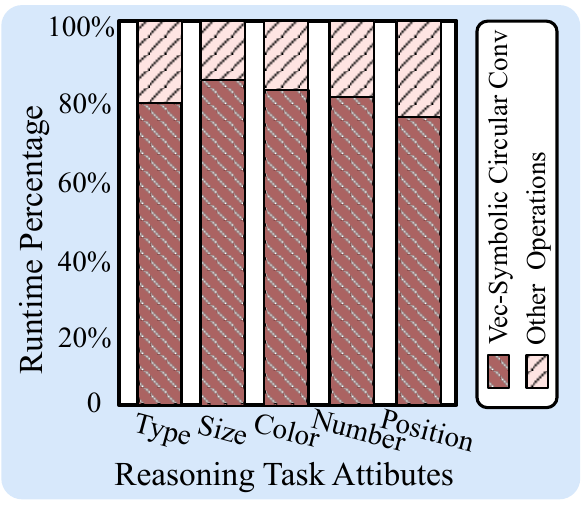}
    \caption{\textbf{Symbolic operation analysis.} \textnormal{Symbolic operations are dominated by vector-symbolic circular convolution and vector-vector multiplication stemming from hypervector representations.}}
    \label{fig:nvsa_backend_operation}
  \end{minipage}
  \vspace{-15pt}
\end{figure}

\begin{table}[t!]
\scriptsize
\centering
\caption{\textbf{Hardware inefficiency analysis.} \textnormal{The compute, memory, and communication characteristics of representative neural and symbolic kernels on CPU/GPU platform.}}
\renewcommand*{\arraystretch}{1.05}
\setlength\tabcolsep{2.3pt}
\resizebox{\linewidth}{!}{%
\begin{tabular}{l|cc|cc}
\hline
\multirow{2}{*}{}        & \multicolumn{2}{c|}{\textbf{Neural Kernel}}                         & \multicolumn{2}{c}{\textbf{Symbolic Kernel}}                \\ \cline{2-5} 
                         & \multicolumn{1}{c|}{sgemm\_nn} & relu\_nn & \multicolumn{1}{c|}{vectorized\_elem} & elementwise \\ \hline
Compute Throughput (\%)  & \multicolumn{1}{c|}{95.1}              & 92.9             & \multicolumn{1}{c|}{3.0}              & 2.3         \\ \hline
ALU Utilization (\%)     & \multicolumn{1}{c|}{90.1}              & 48.3             & \multicolumn{1}{c|}{5.9}              & 4.5         \\ \hline
L1 Cache Throughput (\%) & \multicolumn{1}{c|}{79.7}              & 82.6             & \multicolumn{1}{c|}{28.4}             & 10.8        \\ 
L2 Cache Throughput (\%) & \multicolumn{1}{c|}{19.2}              & 17.5             & \multicolumn{1}{c|}{29.8}             & 22.8        \\ \hline
L1 Cache Hit Rate (\%)   & \multicolumn{1}{c|}{1.6}               & 51.6             & \multicolumn{1}{c|}{29.5}             & 33.3        \\
L2 Cache Hit Rate (\%)   & \multicolumn{1}{c|}{86.8}              & 65.5             & \multicolumn{1}{c|}{48.6}             & 34.3        \\ \hline
DRAM BW Utilization (\%) & \multicolumn{1}{c|}{14.9}              & 24.2             & \multicolumn{1}{c|}{90.9}             & 78.4        \\ \hline
\end{tabular}
}
\label{tab:nsight_profile}
\vspace{-10pt}
\end{table}

\begin{figure*}[b]
\vspace{-0.05in}
\centering\includegraphics[width=2.05\columnwidth]{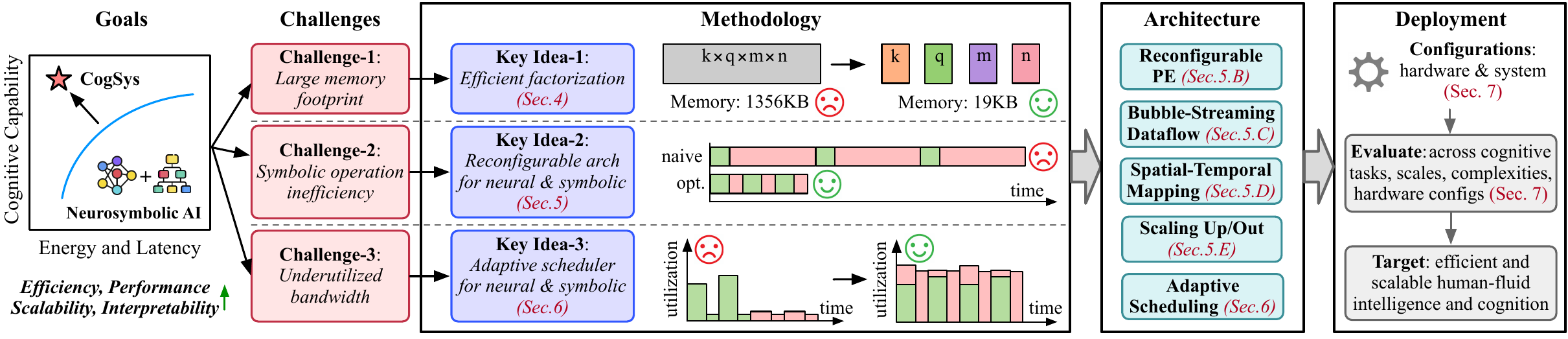}
\vspace{-0.15in}
        \caption{\textbf{\NAME system overview.} \textnormal{\NAME is an algorithm-hardware co-design framework for neurosymbolic AI with the \underline{goal} to achieve efficient and scalable human-fluid intelligence and cognition systems. \NAME addresses the \underline{challenges} of redundant storage, symbolic and vector operation latency bottleneck, sequential processing and hardware underutilization, by proposing \underline{methodologies} including efficient factorization, reconfigurable PE, efficient dataflow, mapping, scalable architecture, multi-level parallelism and scheduler. \NAME is \underline{deployed} across cognitive applications and consistently demonstrates performance-efficiency-accuracy improvements for neurosymbolic systems.}
        }
        \label{fig:leading}
        \vspace{-10pt}
\end{figure*}

\textbf{End-to-end latency scalability.} Fig.~\ref{fig:profiling_latency_e2e}\textcolor{blue}{c} indicates that the neuro vs. symbolic runtime proportion remains relatively stable across various tasks and sizes. For example, when Raven's Progressive Matrices (RPM)~\cite{raven} task size increases from 2$\times$2 to 3$\times$3, the NVSA symbolic modules runtime changes from 91.6\% to 87.4\%, while the total runtime increases by 5.02$\times$ on average across 14 test scenarios, \textit{indicating the scalability bottleneck of neurosymbolic models.} 




\vspace{-5pt}
\subsection{Memory and System Analysis}
\label{subsec:profiling_memory}
\textbf{Memory footprint.}
Fig.~\ref{fig:profiling_latency_e2e}\textcolor{blue}{d} characterizes the memory footprint of neurosymbolic workloads. We can observe that (1) \textit{Neural weights and symbolic codebooks typically consume large storage footprint}, because neural perception enables the expression of more object combinations than vector space dimensions, requiring the codebook to be large enough to ensure vector quasi-orthogonality. (2) Symbolic modules consume large memory due to a large number of vector operations depending on intermediate results and exhaustive search. 


\textbf{System Roofline Analysis.} Fig.~\ref{fig:roofline} employs the roofline model of RTX 2080Ti GPU version to quantify the neurosymbolic workloads. We observe that \textit{symbolic modules are memory-bounded while neuro modules are compute-bounded}. This is mainly due to symbolic operations requiring streaming vector elements, increasing the memory bandwidth pressure and resulting in hardware underutilization (Sec.~\ref{subsec:profiling_symbolic}).

\vspace{-2pt}
\subsection{Symbolic Operation and Inefficiency Analysis}
\label{subsec:profiling_symbolic}



\vspace{-2pt}
\textbf{Symbolic operation analysis.} 
Inspired by the dominated 
symbolic bottleneck, we analyze their detailed operations in Fig.~\ref{fig:nvsa_backend_operation}. We observe that vector-symbolic \textit{circular convolution and vector-vector multiplication dominate symbolic modules}, accounting for 80\% of runtime. In contrast to the shared neural modules, these symbolic operations run sequentially and
separately for each downstream cognition task and underlying rule. These operations typically stem from high-dimensional distributed vectors and are difficult to process efficiently on GPU/TPU. Thus, the challenges of accelerating these vector-symbolic computations will become increasingly important as the cognitive task and feature complexities further grow.

\textbf{Symbolic hardware inefficiency analysis.} To further quantify the reason for hardware inefficiency, we analyze the compute and memory units behavior of representative neuro and symbolic kernels, as listed in Tab.~\ref{tab:nsight_profile}. \emph{The system inefficiencies mainly come from ALU underutilization, low cache hit rate, and massive data movement of symbolic operations.} Symbolic data transfer accounts for half of total latency, where $>$80\% is from host to GPU, while neural kernels exhibit high utilization.








\subsection{Unique Characteristics of Neurosymbolic vs ML Workloads}
\label{subsec:profile_summary}
To summarize, based on above characterization, neurosymbolic AI differs from ML workloads mainly in three aspects:

\textbf{Compute kernels.} Neurosymbolic workloads consist of heterogeneous neural and symbolic kernels. Symbolic operations execute inefficiently on CPU/GPU/TPU with low hardware utilization and cache hit rate, resulting in latency bottleneck.

\textbf{Memory.} Symbolic operations are memory-bounded due to large element streaming for vector-symbolic operations. Symbolic codebooks typically account for large memory footprints and require large intermediate caching during computation. 

\textbf{Dataflow and scalability.} Neurosymbolic workloads exhibit more complex control than CNNs. Symbolic modules typically have irregular dataflow, data dependency, and sequential processing, bringing low parallelism scalability and inefficiency.

\vspace{-2pt}
\subsection{Identified Opportunities for Neurosymbolic Optimization}
\label{subsec:opportunity}


While neurosymbolic AI holds great promise, addressing its inefficiencies is critical for achieving real-time, scalable deployment and ensuring long-term development. To this end, we propose CogSys, an algorithm-hardware co-design framework designed to enhance both reasoning energy efficiency and accuracy in neurosymbolic AI (Fig.~\ref{fig:leading}). \textbf{At the algorithm level}, hardware-friendly codebook optimization reduces memory footprint and latency (Sec.~\ref{sec:algo_opt}). \textbf{At the hardware level}, the architecture and dataflow must be efficient for VSA operations and reconfigurable for neuro/symbolic kernels (Sec.~\ref{sec:accelerator}). \textbf{At the system level}, the architecture must efficiently and adaptively schedule diverse neurosymbolic workloads (Sec.~\ref{sec:sched}). \NAME consistently demonstrates improvements in performance, efficiency, and accuracy across reasoning applications (Sec.~\ref{sec:eval}).

\section{CogSys: Algorithm Optimization}
\label{sec:algo_opt}

This section presents our proposed \NAME algorithm optimizations for efficient and scalable neurosymbolic systems. 
We propose \textit{an efficient vector-symbolic factorization strategy} to reduce the large memory footprint (Sec.~\ref{subsec:algo_factorzation}), and explore the \textit{stochasticity injection} and \textit{low-precision operation} to accelerate neurosymbolic systems (Sec.~\ref{subsec:stochastic}).


\vspace{-2pt}
\subsection{Symbolic Factorization Strategy}
\label{subsec:algo_factorzation}
\textbf{Overall pipeline.} To address the large memory footprint of symbolic codebooks (Sec.~\ref{subsec:profiling_memory}), we propose an efficient factorization strategy. The key idea is to disentangle the large volume of object combination vectors in symbolic knowledge codebook into the small volume of basis attribute vectors, thus lowering computational time and space complexity (Fig.~\ref{fig:factorization}).

Specifically, given an entangled query vector $\boldsymbol{q}$ (e.g., scene representation) generated from neural module and the set of symbolic codebooks $\{X^i\}_{i=1}^{F}$ (each with $M$ possible solutions and $F$ codebooks in total), this creates a combinatorial search and storage involving $M^F$ vectors. Instead of searching along all possible combinations, our factorization method iteratively searches in superposition to find the valid $\boldsymbol{\hat{x}}^i \in X^i$ such that the estimated vector $\boldsymbol{\hat{q}}=\boldsymbol{\hat{x}}^1 \odot \boldsymbol{\hat{x}}^2 \odot \dots \odot \boldsymbol{\hat{x}}^f$ resembles with the highest similarity to the input query $\boldsymbol{q}$. 
By exploiting the quasi-orthogonality of the vectors, our factorization module is able to rapidly search through the various combinations in superposition by iteratively unbinding all but one of the factors from the product vector, and then projecting it into the space of possible solutions of the considered factor that is used for the following reasoning procedure.
In this way, we can replace the original symbolic codebook and greatly reduce its storage.

\textbf{Detailed steps.}
Fig.~\ref{fig:factorization} illustrates our symbolic knowledge codebook factorization strategy, consisting of three steps:

\begin{figure}[t!]
\includegraphics[width=\columnwidth]{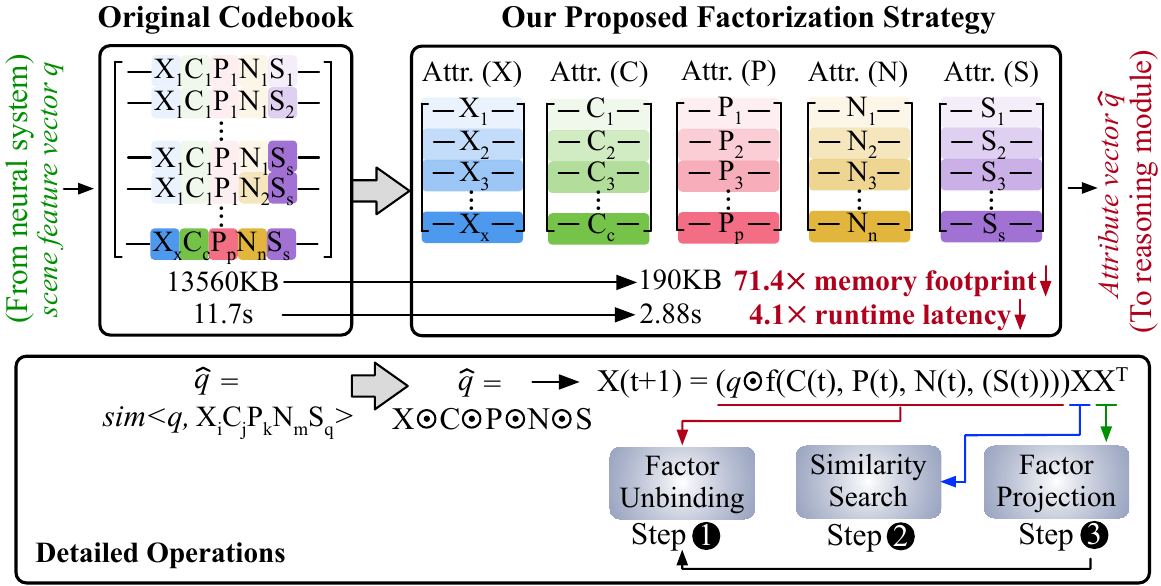}
\vspace{-0.2in}
        \caption{\textbf{Proposed symbolic codebook factorization strategy}. \textnormal{The efficient factorization technique quickly factorizes a product vector in an interactive manner and significantly reduces the memory footprint demand of the symbolic codebook when decomposing query vectors.}}
\label{fig:factorization}
        \vspace{-15pt}
\end{figure}

Step \circled{1}: Factor unbinding via element-wise multiplication ($\oslash$). For a given factor, the unbinding is performed by taking the product vector $\boldsymbol{q}$ and unbinding the contribution of the other factors' latest estimate: $\boldsymbol{\tilde{x}}^{i}(t)=\boldsymbol{q} \oslash \Pi_{f=1}^{F}\boldsymbol{\hat{x}}^{f}(t)$ ($f\ne i$).

Step \circled{2}: Similarity search via matrix–vector multiplication. The similarity vector $\boldsymbol{\alpha}^f(t)$ is calculated for each unbound estimate: $\boldsymbol{\alpha}^f(t)=\boldsymbol{\tilde{x}}^{f}(t) \cdot X^f$, $\forall f \in [1,F]$.

Step \circled{3}: Factor projection via matrix–vector multiplication. The estimates for the factors for the subsequent time step are given by the linear combination of all the codevectors with the similarity vectors acting as weights: $\boldsymbol{\hat{x}}^f(t+1) = \text{sign}(\boldsymbol{\alpha}^f(t) \cdot (X^f)^T)$, $\forall f \in [1,F]$. Repeat Steps \circled{1}-\circled{3} until convergence.

\textbf{Applicable across neurosymbolic workloads.} Our proposed efficient factorization module can apply to various levels of conceptual hierarchy, such as factoring time-varying pixel data of dynamic scenes~\cite{anderson2020high}, factoring sentence structure into roles and fillers~\cite{liu2022verb}, and cognitive analogical reasoning~\cite{hersche2023neuro}. Essentially, given its ubiquitous applicability in perception and cognitive reasoning, we envision it being a core component in future large-scale neurosymbolic cognitive systems.

\vspace{-2pt}
\subsection{Stochasticity and Low-Precision Operation}
\label{subsec:stochastic}
\textbf{Factorization optimization via stochasticity.} To reduce the required number of iterations of factorization, we propose to apply additive Gaussian noise. We observe that injecting stochasticity in both Step \circled{2} similarity and Step \circled{3} projection operations helps the factorization process escape limit cycles, allowing it to explore a much larger solution space and achieve faster convergence (Tab.~\ref{tab:eval_acc}).


\textbf{Operator precision optimization.} To further reduce the memory footprint, we apply quantization techniques to the workloads. Specifically, we apply 8-bit floating-point and integer arithmetic for both neuro and symbolic computations with fine-tuning to maintain reasoning accuracy (Tab.~\ref{tab:eval_quant}).

\subsection{Algorithm Optimizations Discussion}
\begin{table}[t!]
\centering
\caption{\textbf{Algorithm optimization impact.} \textnormal{Factorization, stochasticity, and quantization impact accuracy, latency, and memory.}}
\renewcommand*{\arraystretch}{1.2}
\setlength\tabcolsep{2.0pt}
\resizebox{\linewidth}{!}{%
\begin{tabular}{c|c|c|c}
\hline
              & \textbf{Accuracy} (higher is better) & \textbf{Latency} (lower is better) & \textbf{Memory} (lower is better) \\ \hline
Factorization & Increase                  & Reduce                  & Reduce                 \\ \hline
Stochasticity & Increase / Reduce           & Reduce                  & No Impact              \\ \hline
Low-Precision  & Reduce                    & Reduce                  & Reduce                 \\ \hline
\end{tabular}
}
\label{tab:algo_opt_comp}
\vspace{-15pt}
\end{table}

\textbf{Impact on accuracy, latency, and memory.} The factorization, stochasticity, and quantization optimizations impact accuracy, latency, and memory requirements to varying extents. As shown in Tab.~\ref{tab:algo_opt_comp}, accuracy improves with factorization (due to precise attribute extraction, aiding downstream symbolic reasoning) and with stochasticity optimizations (due to improved convergence). However, quantization results in accuracy decreases due to data imprecision. Designers can balance speed and accuracy by tuning factorization convergence threshold.

\section{CogSys: Hardware Architecture}
\label{sec:accelerator}

This section presents \NAME architecture, the \emph{first} hardware to enable efficient and scalable neurosymbolic processing. \NAME architecture features \textit{reconfigurable neuro/symbolic processing elements (nsPE)} (Sec.~\ref{subsec:hw_circular_conv}), \textit{bubble streaming (BS) dataflow} (Sec.~\ref{subsec:HW_dataflow}),
\textit{adaptive spatial-temporal (ST) mapping} (Sec.~\ref{subsec:folding}) with scalable array architecture (Sec.~\ref{subsec:hardware_systolic_array}) and customized SIMD units (Sec.~\ref{subsec:simd}) for neurosymbolic processing.

\begin{figure*}[t!]
  \centering
  \begin{minipage}[t]{1.15\columnwidth}
    \centering
    \includegraphics[width=.95\textwidth]{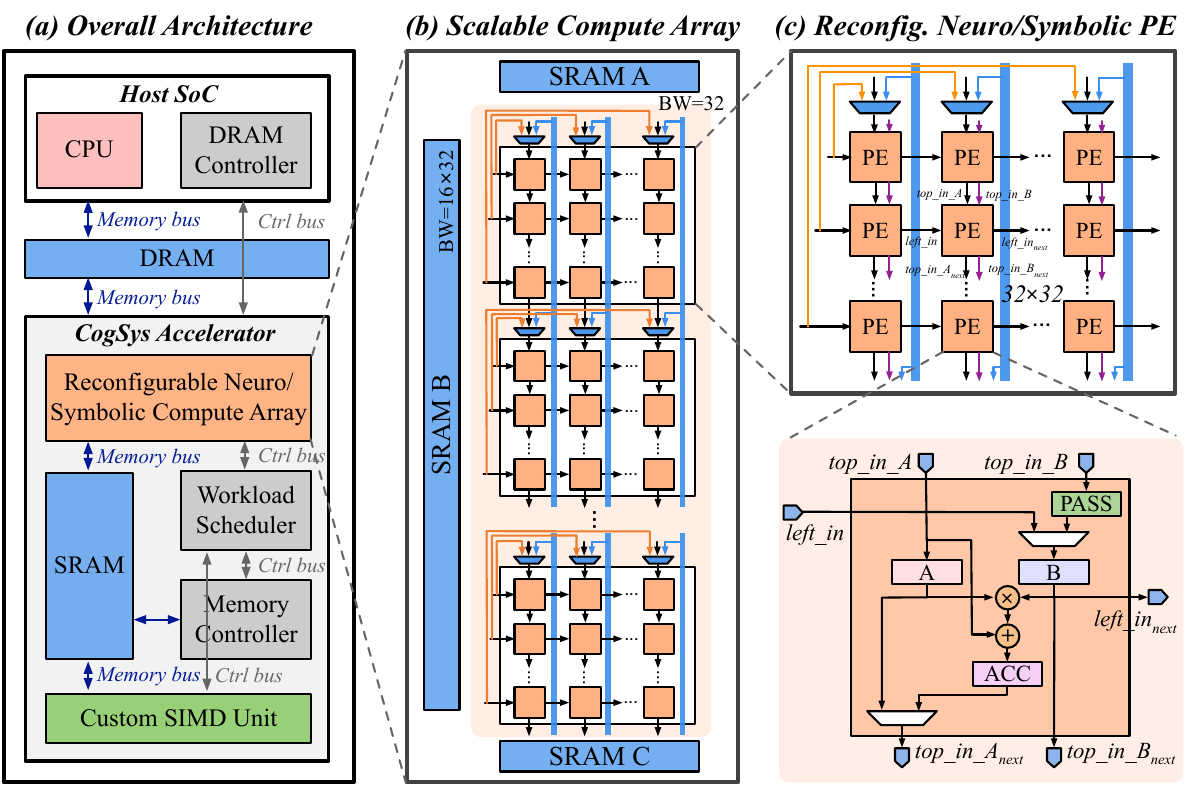}
    \caption{\textbf{Architecture overview.}
        \textnormal{\NAME system includes DRAM, host System-on-Chip (SoC), and \NAME accelerator that consists of five major components: reconfigurable and scalable neuro/symbolic compute arrays, custom SIMD units, double-buffered SRAMs, workload scheduler, and memory controller.}}
    \label{fig:arch}
  \end{minipage}
  \hfill
  \begin{minipage}[t]{.83\columnwidth}
    \centering
    \includegraphics[width=.95\textwidth]{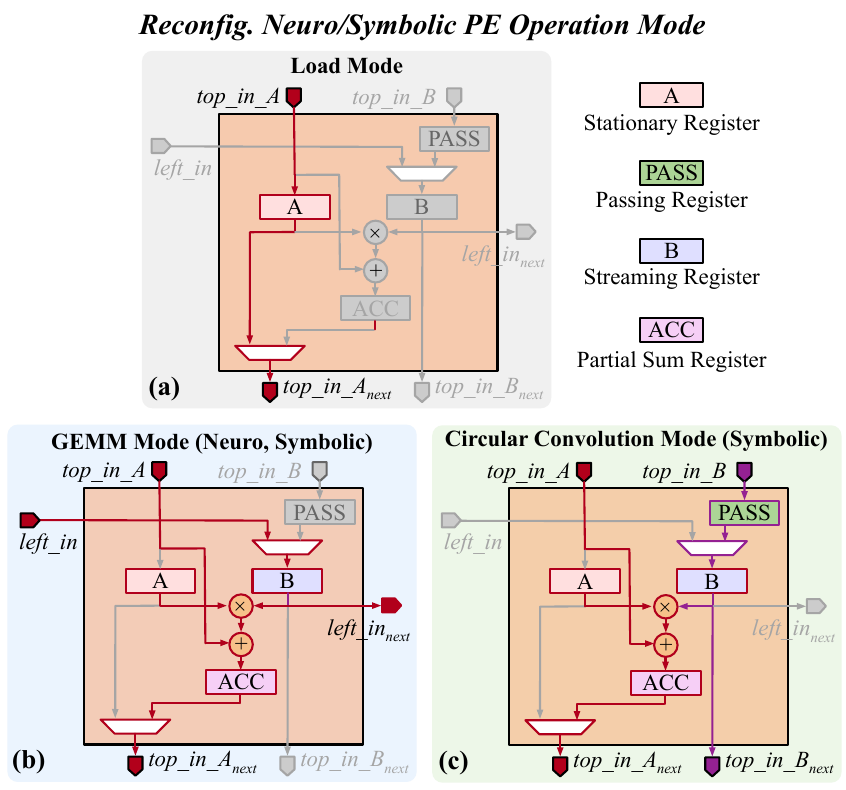}
    \caption{\textbf{Reconfigurable neuro/symbolic PE (\textit{nsPE}).} \textnormal{Each \textit{nsPE}} includes four registers and supports three modes (load, neuro, symbolic) that provide reconfigurable support for neurosymbolic operations.}
    \label{fig:reconfig_pe}
  \end{minipage}
  \vspace{-10pt}
\end{figure*}


\begin{table}[t!]
\centering
\caption{\textbf{Comparison between \NAME with other accelerators.} \textnormal{\NAME is the first to enable efficient and scalable vector-symbolic circular convolution (CircConv) and neurosymbolic models.
}}
\renewcommand*{\arraystretch}{1.2}
\setlength\tabcolsep{1.8pt}
\resizebox{\linewidth}{!}{%
\begin{tabular}{c|c|c|c|c|c|c}
\hline
\textbf{Accelerators} & \begin{tabular}[c]{@{}c@{}}\textbf{Reconfi-} \\ \textbf{gurable}\end{tabular} & \begin{tabular}[c]{@{}c@{}}\textbf{Scale-up/}\\ \textbf{Scale-out}\end{tabular} & \begin{tabular}[c]{@{}c@{}@{}}\textbf{Memory Footprint for}\\ \textbf{Single CircConv} \\ \textbf{(Vector Dimension = $\mathbf{d}$)}\end{tabular} & \begin{tabular}[c]{@{}c@{}@{}}\textbf{CWP$^*$} \\ \textbf{Support for} \\ \textbf{CircConv} \end{tabular}  & \begin{tabular}[c]{@{}c@{}@{}}\textbf{ScWP$^{**}$} \\ \textbf{Support for} \\ \textbf{CircConv} \end{tabular} & \begin{tabular}[c]{@{}c@{}}\textbf{Neuro-}\\ \textbf{symbolic AI}\\\textbf{ Support}\end{tabular} 

\\ \hline
Eyeriss~\cite{chen2016eyeriss} & No                                              &   Scale-up                                            &     No Support                                          &   No & No                                            &    No                                                  \\ \hline
Neuro Cube~\cite{kim2016neurocube} &    No                                           &   Scale-out                                            &  $O(d^2)$, GEMV                                             &    No & Yes                                          &      No                                                \\ \hline
Brainwave~\cite{chung2018serving} &   No                                            &        Scale-out                                       &   $O(d^2)$, GEMV                                            &    No                                           &  Yes &   No                                                \\ \hline
SARA \cite{samajdar2022self} &     No                                          &        Both                                       &       $O(d^2)$, GEMV                                        &      No     & Yes                                    &           No                                           \\ \hline
TPU~\cite{jouppi2020domain} &     No                                          &    Scale-out                                           &    $O(d^2)$, GEMV                                           &     No   & Yes                                       &    No                                                  \\ \hline
SIMBA~\cite{shao2019simba} &     No                                          &    Scale-out                                           &    $O(d^2)$, GEMV                                           &     No  & Yes                                        &    No                                                  \\ \hline
MTIA~\cite{firoozshahian2023mtia} &    No                                           &   Scale-out                                            &  $O(d^2)$, GEMV                                             &    No & Yes                                           &      No                                                \\ \hline
\textbf{\NAME (Ours)} &    \textbf{Yes}                                           &          \textbf{Both}                                     &    \textbf{$\mathbf{O(d)}$, BS Dataflow}                                          &     \textbf{Yes}  & \textbf{Yes}                                        &  \textbf{Yes}                                                    \\ \hline
\end{tabular}
}

\label{tab:accelerator_comparison}
\begin{threeparttable}
\vspace{2pt}
   \scriptsize{*Column-Wise Parallelism (within a systolic cell); **Systolic cell-Wise Parallelism}

\end{threeparttable}
\vspace{-2em}
\end{table}






\vspace{-3pt}
\subsection{Overview of Proposed \NAME Architecture}
\label{subsec:hardware_overview}
Neurosymbolic workloads feature much greater heterogeneity in compute kernels than DNNs, leading to an increasing divergence with the current hardware that focuses on GEMMs and convolutions. As illustrated in Tab.~\ref{tab:accelerator_comparison}, \NAME is proposed, for the first time, to support neurosymbolic workloads and achieve efficient and scalable implementation of symbolic operations.

Aiming to design a complete neurosymbolic acceleration system, our design includes DRAM, a host SoC, and \NAME accelerator consisting of five major components: reconfigurable neuro/symbolic compute array,
SIMD unit, double-buffered SRAM, adaptive scheduler and memory controller (Fig.~\ref{fig:arch}). During the reasoning procedure, the host SoC streams task in, and then the reconfigurable arrays perform neuro (e.g., GEMMs/convolutions) and vector-symbolic operations (e.g., circular convolution), while the SIMD units accelerate element- and vector-wise operations with multi-level parallelism and adaptive workload scheduling. \emph{It is worth noting that monolithic systolic array (TPU-like) is extremely inefficient for symbolic workloads (Sec.~\ref{subsec:HW_dataflow}), while \NAME provides reconfigurable support for neural and symbolic kernels and demonstrates $>$75$\times$ speedup with only $<$5\% area overhead over systolic array architecture.}

\subsection{Reconfigurable Neuro/Symbolic Processing Element}
\label{subsec:hw_circular_conv}



\textbf{Reconfigurable neuro/symbolic PE (\textit{nsPE}).}
Instead of having separate PEs for neuro and symbolic operations that incur large area overhead, we propose \textit{nsPE} micro-architecture that provides reconfigurable support to both neuro and symbolic operations (Fig.~\ref{fig:reconfig_pe}). Each \textit{nsPE} consists of four registers (stationary, passing, streaming, and partial sum registers) and supports three operation modes (load, GEMM, and circular convolution). 
\underline{During load mode}, the input vectors $\boldsymbol{A}$ (weights of GEMM) are passed into the stationary register using `top\_in\_A' links.
Reconfigurability is achieved by selecting input $\boldsymbol{B}$ either from `left\_in' link (GEMM mode) or the passing register (circular convolution mode). \underline{During GEMM}
\underline{mode}, the \textit{nsPE} operates as TPU-like architecture for efficient GEMM and convolution. Inputs are streamed from left to right using `left\_in' links. \underline{During circular convolution mode}, input vector $\boldsymbol{B}$ is streamed from top to bottom using `top\_in\_B' links with a bubble via passing register (Sec.~\ref{subsec:HW_dataflow}), facilitating the temporal reuse of the streaming input for efficient vector-symbolic circular convolution operation. The reconfigurable \textit{nsPE} can also support efficient circular correlation by reversing stationary vector $\boldsymbol{A}$. During both GEMM and circular convolution modes, partial products are reduced from top to bottom with `top\_in\_A' links.
 

\subsection{Efficient Bubble Streaming Dataflow}
\label{subsec:HW_dataflow}

\begin{figure*}[t!]
\centering
\vspace{-0.02in}
\includegraphics[width=\textwidth]{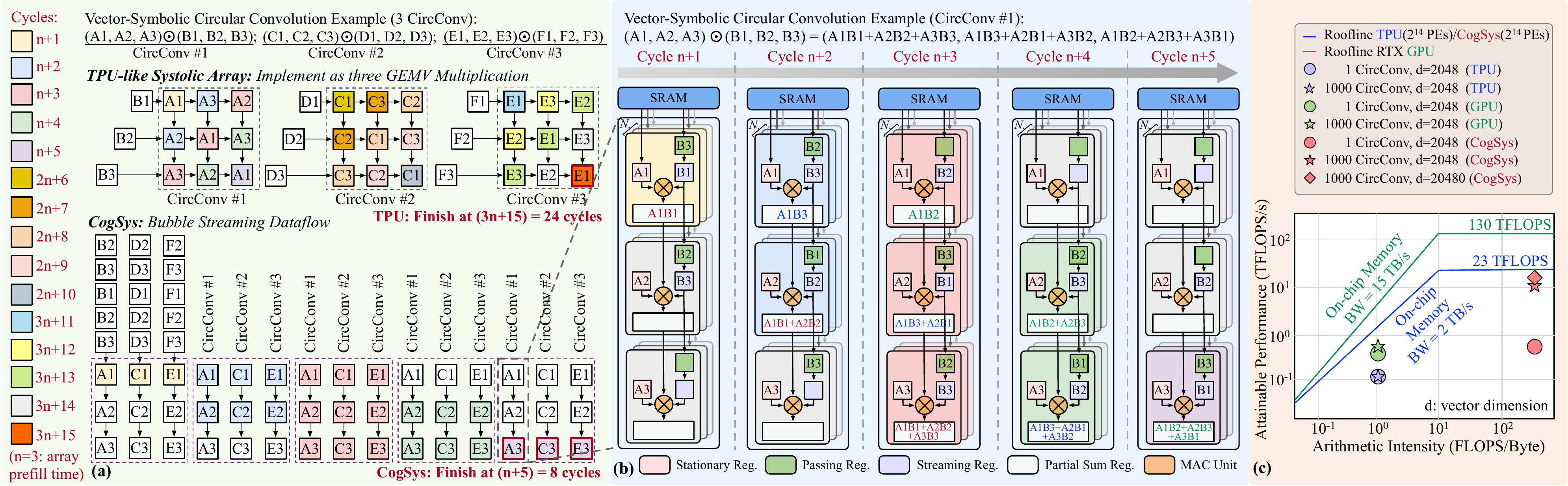}
\vspace{-0.15in}
        \caption{\textbf{Efficient bubble streaming (\textit{BS}) dataflow, roofline analysis, CogSys/TPU/GPU comparison.} \textbf{(a)} Compute cycle and mapping comparison of TPU-like systolic array dataflow and \NAME \textit{BS} dataflow under multiple circular convolutions. \textbf{(b) }\textnormal{\textit{BS} dataflow showing circular convolution of two vectors $\boldsymbol{A}$ and $\boldsymbol{B}$ ($d$=3) in a 3$\times$1 \NAME \textit{nsPE} array. \textbf{(c)} Roofline comparison of circular convolution implemented as \textit{BS} dataflow (compute-bound) on \NAME ($2^{14}$ PEs) against implemented as GEMV in TPU systolic cell ($2^{14}$ PEs) and GPU (memory-bound).}
        }
        \label{fig:circular_convolution}
        \vspace{-15pt}
\end{figure*}



\textbf{Inefficiency of TPU-like systolic array.} 
TPU-like systolic array (SA) exhibits high memory footprint and low parallelism for symbolic circular convolution operations. Fig.~\ref{fig:circular_convolution}\textcolor{blue}{a} shows a scenario of three circular convolutions. TPU-like systolic cell implements them as general matrix-vector (GEMV) multiplication where matrices contain circularly shifted stationary vectors with the matrix memory footprint of O($d^2$). Additionally, TPU-like SA is incapable of parallelizing multiple GEMV on a systolic cell and need to process them sequentially.

\textbf{Bubble streaming (\textit{BS}) dataflow.} To efficiently support symbolic operations in \textit{nsPE} array, we propose \textit{BS} dataflow for circular convolution (Sec.~\ref{subsec:vsa_background}) and circular correlation (opposite direction circular shift) which is the vector-symbolic bottleneck.
Fig.~\ref{fig:circular_convolution}\textcolor{blue}{b} presents an example of \textit{BS} dataflow performing circular convolution of two vectors $\boldsymbol{A}$ and $\boldsymbol{B}$ ($d$=3) on a 3$\times$1 \textit{nsPE} array. In \textit{BS} dataflow, vector $\boldsymbol{B}$ is streamed from one \textit{nsPE} to another through bubbles (passing registers) while vector $\boldsymbol{A}$ is held in stationary registers. The \textit{BS} dataflow enables a passing register to temporarily store the streaming input for a cycle before it moves to the streaming register. This value is transferred to the passing register of the next \textit{nsPE} in the following cycle. The MAC unit processes the data from stationary and streaming registers, adding it to the partial product. The procedure is repeated until final outputs.

\textbf{Improved arithmetic intensity of \textit{BS} dataflow.} The \textit{BS} dataflow achieves higher arithmetic intensity than GEMV in GPU/TPU-like systolic cells, as illustrated in roofline analysis (Fig.~\ref{fig:circular_convolution}\textcolor{blue}{c}). This efficiency mainly comes from reduced memory footprint and increased parallelism (Tab.~\ref{tab:accelerator_comparison}). 
(1) Linear memory footprint: The bubble enables circularly shift in vector $\boldsymbol{B}$ ($O(d)$) and alleviates the overhead of creating and fetching matrix ($O(d^2)$) of circularly shifted $\boldsymbol{B}$ as TPU-like systolic cell, reducing footprint by $d \times$.
(2) Column-wise parallelism (CWP):
The \textit{BS} dataflow enables each column of a systolic cell to execute a circular convolution, thus multiple circular convolutions can be parallelized over multiple columns, which is not possible for GEMV in a TPU-like systolic cell.  
The low arithmetic intensity and memory-bound operations make GPUs inefficient for vector-symbolic circular convolution. For circular convolution of two $d$-dimensional vectors, the GPU arithmetic intensity is $d\times(d+d-1)/(d\times d+d\times1+d\times1)$, while CogSys arithmetic intensity is $d\times(d+d-1)/(d\times1+d\times1+d\times1)$. As in Fig.~\ref{fig:circular_convolution}\textcolor{blue}{c}, CogSys achieves peak performance when fully utilized, while GPU suffers from memory-bound. Additionally, GPUs require extra computations to handle the index calculations for the circularly shifted vector.

\textbf{Cycle analysis of \textit{BS} dataflow.} Assuming the case of \textit{nsPE} array size $M =$ input vector size $d$ for vector-symbolic circular convolution, streaming the stationary input would take $d$ cycles followed by input vectors taking $2d$ cycles to reach the final \textit{nsPE}. Meanwhile, partial sums are aggregated along the array for the first output, followed by the remaining ($d-1$) outputs where each is generated per cycle. Thus, the end-to-end latency for vector-symbolic circular convolution of two $d$-dimensional vectors in a 1-D \textit{nsPE} compute array is $(4d-1)$ cycles. In the case where $d\neq$ M, latency $T$ will be $(3M+d-1)$ cycles where $M$ for loading stationary vector, $2M$ for streaming vector reach final \textit{nsPE}, and $(d-1)$ cycles for remaining outputs. 

\vspace{-2pt}
\subsection{Adaptive Spatial and Temporal Mapping Strategy}
\label{subsec:folding}
\begin{figure}
\begin{minipage}[t!]{\linewidth}
    \centering
    \includegraphics[width=\columnwidth]{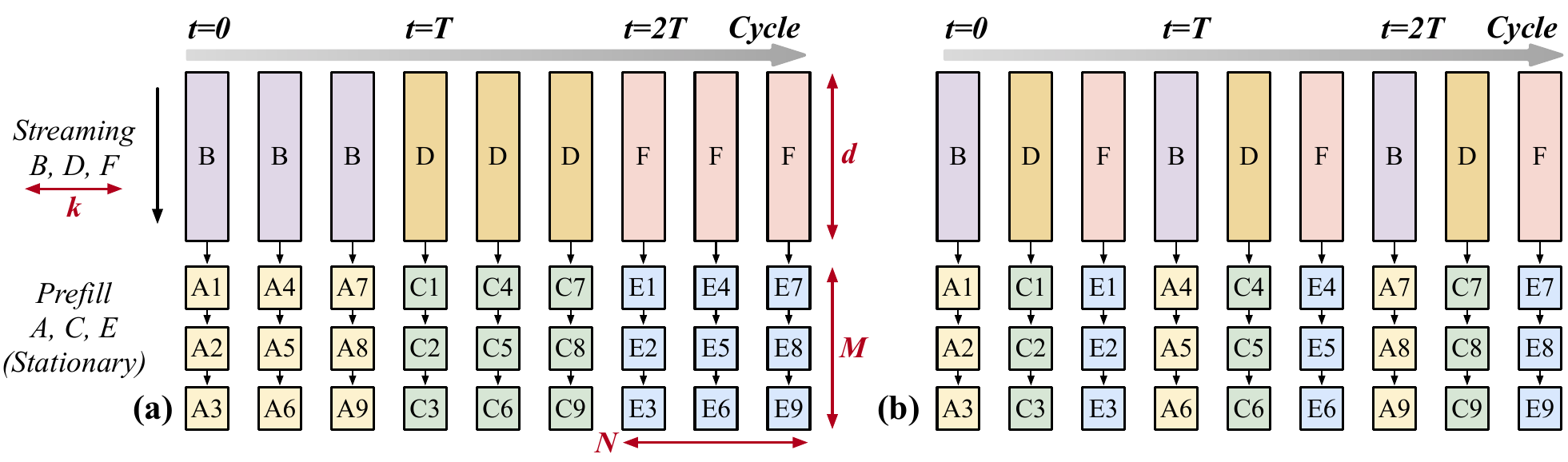}
\end{minipage}%
 \vspace{0.05in}
\begin{minipage}[b]{\columnwidth}
\renewcommand*{\arraystretch}{1.2}
\setlength\tabcolsep{2pt}
\resizebox{\linewidth}{!}{%
\begin{tabular}{c|c|c}
\hline
& \textbf{(a) Spatial Mapping} & \textbf{(b) Temporal Mapping}\\\hline
\textbf{Latency} & $k \times \lceil d/(N \times M) \rceil \times T$
& $\lceil k/N \rceil \times \lceil d/M \rceil \times T$ \\
\hline
\textbf{Mem Reads under full util.} & $2d$ per T cycles & $(d+M)\times N$ per T cycles \\
\hline
\end{tabular}
}
\caption{\textbf{Adaptive spatial-temporal (\textit{ST}) mapping.} The \textit{ST} mapping under \textit{BS} dataflow. \NAME adaptively chooses the mapping scheme based on $(N, M, k, d)$ workload and hardware configurations.}
\vspace{-15pt}
\label{fig:spatial_temporal_map}
\end{minipage}
\end{figure}

\textbf{Spatial-temporal (\textit{ST}) mapping flexibility.} To efficiently support the various dimensions of vector-symbolic operations, we propose \textit{ST} mapping featuring \underline{spatial mapping mode} and \underline{temporal mapping mode} (Fig.~\ref{fig:spatial_temporal_map}). The \textit{nsPE} array is easily expanded to $N$ arrays.
Spatial mapping, by parallelizing a single circular convolution into folds, reduces memory reads compared to temporal mapping, especially with many folds. Taking $N$ arrays ($M$ PEs each) as an example, spatial mapping requires $B_S$=$2d$ memory reads per $T$ cycles for $d$-dimensional vectors. while temporal mapping requires loading $B_T$=($d$+$M$)$\times$$N$ elements per $T$ cycles. Given neurosymbolic workloads typically have $d$$>$1000, the bandwidth requirement (memory reads per $T$ cycles) is reduced by $(N/2)\times$ via spatial mapping. Temporal mapping, on the other hand, outperforms spatial mapping when $d$$<$$M$ by enabling the parallelization of multiple convolutions. For $k$ vector-symbolic circular convolutions, temporal folding takes 
 $C_T$=$\lceil k/N \rceil$$\times$$\lceil d/M \rceil$$\times$$T$ cycles, while spatial folding takes $C_S$=$k$$\times$$\lceil d/(N \times M) \rceil$$\times$$T$ cycles.

\begin{figure*}[t]
\centering
\vspace{-0.02in}
\includegraphics[width=\textwidth]{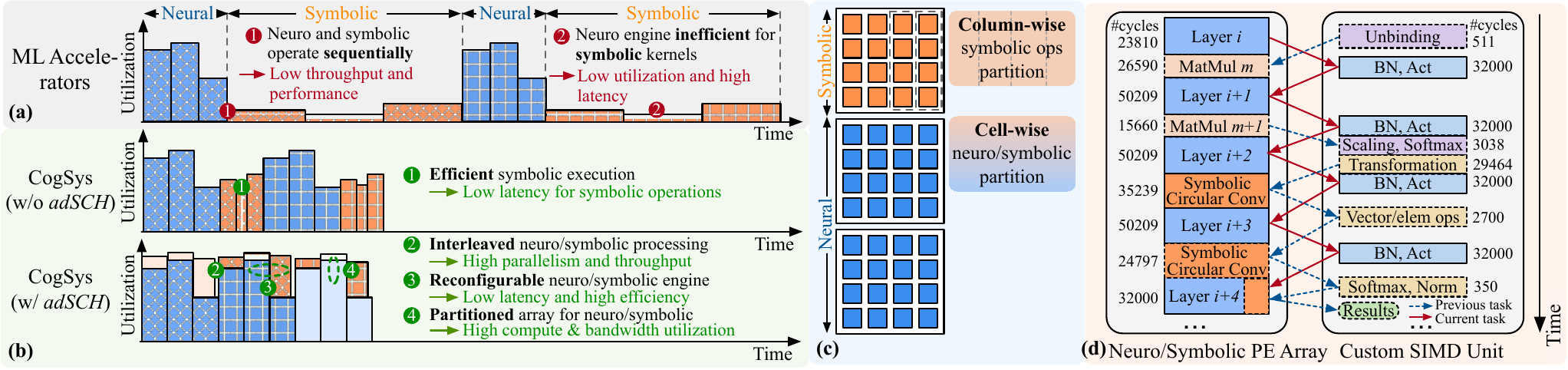}
\vspace{-0.2in}
        \caption{\textbf{Adaptive workload-aware scheduling (\textit{adSCH}) strategy.} \textnormal{\textbf{(a)} Neurosymbolic system-level challenges. The \textit{adSCH} scheme enables \textbf{(b)} interleaved neural/symbolic processing and \textbf{(c)} cell-wise partition across \NAME arrays with multi-level parallelism. \textbf{(d)} An \textit{adSCH} example on NVSA algorithm. The scheduling method ensures \NAME is adaptable and scalable across neurosymbolic workloads and tasks.}}
        \label{fig:scheduling}
        \vspace{-15pt}
\end{figure*}

 To efficiently process symbolic operations and balance between bandwidth and latency, we conduct an adaptive search between spatial and temporal mapping based on workloads and \NAME configurations. For example, For $N$=32 and $d$=1024 in NVSA ($k$=210) and LVRF ($k$=2575) workloads, \NAME opts for temporal mapping with 32 parallel circular convolutions.

\vspace{-3pt}
\subsection{Adaptive Scale-Up and Scale-Out Strategy}
\label{subsec:hardware_systolic_array}
\textbf{Scale-up and scale-out flexibility.}
To enhance design utilization and scalability, \NAME proposes to operate as a combination of scale-up and scaled-out reconfigurable arrays,
with the support of \underline{systolic cell-wise parallelism (ScWP)} and \underline{column-wise parallelism (CWP)} for circular convolutions. The ($N$=32, $M$=512) configuration is constructed from 16 32$\times$32 cells by configuring the muxes to choose among five schemes, i.e., scale-up GEMM, scale-out GEMM, scale-up Conv, scale-out Conv, and scale-out GEMM+Conv, enabled by the systolic cell-wise heterogeneous partitioning scheme in Sec.~\ref{subsec:scheduling}. For GEMM, the scale-out scheme enables higher utilization and ScWP. For symbolic circular convolution, the scaled-out scheme enables ScWP and CWP for low-dimensional vectors.

\textbf{Design space exploration.} We search scale-out/scale-up schemes based on workloads and \NAME configurations to increase utilization and parallelism. For instance, the 16 32$\times$32 scaled-out cells achieve 91.26\% utilization, with 10.71$\times$ and 7.83$\times$ speedup over one 128$\times$128 scaled-up and four 64$\times$64 scaled-out cells, respectively, for NVSA and LVRF neural modules. For vector-symbolic operations, \NAME chooses a scale-up scheme for NVSA and LVRF (high-dimensional vector processing, $d$=1024) and a scale-out scheme for MIMONet (low-dimensional vector processing, $d$=64).

\vspace{-3pt}
\subsection{Double-buffered Memory and Custom SIMD Unit} 
\label{subsec:simd}
\textbf{Double-buffered memory.} \NAME array is backed by three double-buffered SRAMs (Fig.~\ref{fig:arch}\textcolor{blue}{b}). The double-buffered memory is effective in reducing off-chip accesses and stalls due to loads and stores to reduce latency. SRAM A is common for all cells to utilize weights temporal reuse while SRAM B is distributed across cells. Through design space exploration, \NAME opts for 256kB for SRAM A and 4MB for SRAM B.

\textbf{Custom SIMD units.} \NAME employs a custom SIMD unit to execute vector reductions and element-wise operations (Fig.~\ref{fig:arch}\textcolor{blue}{a}). The SIMD unit efficiently handles data transfer between the \NAME array output and input SRAM, enabling the array to seamlessly access data for subsequent operations. The SIMD unit is comprised of multiple PEs, each designed with compact logic circuits (i.e., sum, mult/div, exp/log/tanh, norm, softmax, etc) to perform vector operations on quantized data. The adaptive workload-aware scheduling (Sec.~\ref{sec:sched}) scheme schedules workloads across \NAME array and SIMD units to balance the runtime of neural and symbolic operations.

\begin{table}[t!]
\centering
\caption{\textbf{Design choice discussion.} \textnormal{Area, latency, energy, and utilization comparison with reconfigurable and heterogeneous PEs.}}
\renewcommand*{\arraystretch}{1.1}
\setlength\tabcolsep{2.0pt}
\resizebox{\linewidth}{!}{%
\begin{tabular}{c|c|c|c|c|c}
\hline
                           & \textbf{Configuration}                                                                             & \textbf{Area}  & \textbf{Latency} & \textbf{Energy}                      & \textbf{Utilization} \\ \hline
Reconfigurable PE (CogSys) & \begin{tabular}[c]{@{}c@{}}16$\times$32$\times$32 reconfigurable\\ neuro/symbolic PE\end{tabular}                                               & 1$\times$   & 1$\times$      & 1$\times$                          & 90\%                \\ \hline
Heterogeneous PE           & \begin{tabular}[c]{@{}c@{}}16$\times$32$\times$32 neuro PE\\ 16$\times$32$\times$32 symbolic PE\end{tabular} & 1.96$\times$ & 1$\times$      & 1.3$\times$ & 45\%                \\ \hline
Heterogeneous PE           & \begin{tabular}[c]{@{}c@{}}8$\times$32$\times$32 neuro PE\\ 8$\times$32$\times$32 symbolic PE\end{tabular}   & 0.98$\times$ & 2$\times$      & 1.3$\times$  & 45\%                \\ \hline
\end{tabular}
}
\label{tab:reconfig_heter}
\vspace{-15pt}
\end{table}

\vspace{-2pt}
\subsection{Design Choices Discussion}
\textbf{Reconfigurable or heterogeneous PE.} While specialized PEs for neural and symbolic kernels may appear more efficient for simultaneous processing, our early-phase design space exploration reveals that using separate PEs for GEMM and circular convolution leads to hardware underutilization and increased area overhead due to the sequential execution of neural and symbolic kernels, as well as the varying proportions of these operations across different workloads. When comparing designs with the same chip size, specialized PEs result in longer latency, as they provide fewer effective compute units for either neural or symbolic tasks, as summarized in Tab.~\ref{tab:reconfig_heter}. We thus opt for the reconfigurable PE approach, which offers lower area overhead and higher hardware utilization, making it suitable for both neuro-heavy and symbolic-heavy workloads.

\section{CogSys: Scheduling Strategy}
\label{sec:sched}

This section presents \NAME adaptive workload-aware scheduling strategy. 
We first identify the system-level challenges of neurosymbolic workloads (Sec.~\ref{subsec:system_challenge}), and then introduce \NAME scheduling scheme (Sec.~\ref{subsec:scheduling}) and further discuss its scalability to other workloads and tasks (Sec.~\ref{subsec:scalability}).

\vspace{-2pt}
\subsection{Neurosymbolic System-Level Challenges}
\label{subsec:system_challenge}
We identify two main neurosymbolic system-level challenges (Fig.~\ref{fig:scheduling}\textcolor{blue}{a}): \underline{First}, the sequential execution and frequent interactions of neural and symbolic components results in long latency and low system throughput. \underline{Second}, the heterogeneous neural and symbolic kernels result in low compute array utilization and efficiency of ML accelerator.

\subsection{Adaptive Workload-Aware Scheduling (adSCH) Strategy}
\label{subsec:scheduling}
\textbf{Adaptive scheduling (\textit{adSCH}) strategy.}
To solve the system-level challenges, \NAME features an \textit{adSCH} scheme and greatly improves hardware utilization and performance. 
\underline{(1) Interleaved neural/symbolic processing.} Despite the dependencies in neural and symbolic tasks, symbolic operations of other tasks can be interleaved within neural layer of current task via reconfigurable neuro/symbolic PE arrays (Fig.~\ref{fig:scheduling}\textcolor{blue}{b}). 
\underline{(2) Adaptive neuro/symbolic array partition strategy.} We propose to adaptively allocate \NAME cells to various neural and symbolic kernels (cell-wise partition), and allocate symbolic cell columns to parallel circular convolution operations (column-wise partition) (Fig.~\ref{fig:scheduling}\textcolor{blue}{c}). This partition strategy is effective in handling both neural- and symbolic-intensive workloads and promotes parallelism and hardware utilization.

\textbf{Scheduling Implementation.} CogSys workload-aware scheduling is performed offline by software. Since the model architecture, size, and data are known prior to execution, the host CPU precomputes the mapping of operations and array configurations, which are then offloaded to CogSys. This ensures optimal or near-optimal scheduling with zero runtime latency. The scheduling process uses a greedy search algorithm: (1) Generate an operation graph based on operation type, size, dependencies, and number of iterations; (2) Assign ready operations (not blocked by dependencies) to newly available cells, with runtime estimated analytically; (3) Maximize utilization by prioritizing neural tasks for larger cell blocks and symbolic tasks for smaller ones. Since the search only considers available blocks within the 16 array cells and ready tasks, the search space is limited to $<$$O(10^3)$ per time step, resulting in minimal offline overhead and no runtime overhead.

\textbf{Adaptive scheduling example.} Fig.~\ref{fig:scheduling}\textcolor{blue}{d} presents a detailed example of \textit{adSCH} scheme with operations and cycle numbers in a NVSA segment~\cite{hersche2023neuro}. \NAME reconfigurable array schedules neural (convolutions, GEMMs) and symbolic (circular convolutions), while element-wise operations are offloaded to SIMD units. To mitigate underutilization, \NAME executes VSA-based codebook and symbolic kernels of the previous batch on idle hardware pieces during neural layers of the current batch, thus eliminating symbolic bottleneck. Particularly, \emph{multi-level parallelism} is adopted to process different symbolic rules and attributes to further improve efficiency.


\begin{table}[t!]
\centering
\caption{\textbf{Baseline.} \textnormal{The specifications of hardware baseline. 
}}
\renewcommand*{\arraystretch}{1.4}
\setlength\tabcolsep{1.9pt}
\resizebox{\linewidth}{!}{%
\begin{tabular}{c|cccc|c||c|ccc|c}
\hline
\multirow{2}{*}{HW} & \multicolumn{4}{c|}{General Purpose Processor/SoC}  & \multirow{2}{*}{\begin{tabular}[c]{@{}c@{}}CogSys\\ (Ours)\end{tabular}}                                                                                                                             & \multirow{2}{*}{HW} & \multicolumn{3}{c|}{ML Accelerator}                                            & \multirow{2}{*}{\begin{tabular}[c]{@{}c@{}}CogSys\\ (Ours)\end{tabular}} \\ \cline{2-5} \cline{8-10}
                          & \multicolumn{1}{c|}{CPU Xeon}                  & \multicolumn{1}{c|}{RTX GPU}                   & \multicolumn{1}{c|}{TX2}               &   NX            &   &                        & \multicolumn{1}{c|}{TPU-like}  & \multicolumn{1}{c|}{MTIA-like} & Gemmini-like &                                                                          \\ \hline
\multirow{2}{*}{Power}    & \multicolumn{1}{c|}{\multirow{2}{*}{145W}} & \multicolumn{1}{c|}{\multirow{2}{*}{250W}} & \multicolumn{1}{c|}{\multirow{2}{*}{15W}} & \multirow{2}{*}{20W} & \multicolumn{1}{c||}{\multirow{2}{*}{1.48W}} & SRAM                      & \multicolumn{1}{c|}{4.5MB}     & \multicolumn{1}{c|}{4.5MB}     & 4.5MB        & 4.5MB                                                                    \\ \cline{7-11} 
                          & \multicolumn{1}{c|}{}                     & \multicolumn{1}{c|}{}                     & \multicolumn{1}{c|}{}                      &     &                  & \#PE                      & \multicolumn{1}{c|}{1 128$\times$128} & \multicolumn{1}{c|}{16 32$\times$32}  & 64 16$\times$16     & 16 32$\times$32                                                                 \\ \hline
\end{tabular}
}
\label{tab:eval_hw_baseline}
\vspace{-15pt}
\end{table}

\subsection{Scalability and Variability Support}
\label{subsec:scalability}
\textbf{Scalable across neurosymbolic workloads and cognition tasks.}
The \textit{adSCH} technique enables \NAME to be easily reconfigured across \underline{(1) neurosymbolic workloads} (e.g., NVSA, MIMONets, LVRF, etc) and \underline{(2) cognitive tasks} such as procedurally generated matrices (PGM)~\cite{barrett2018measuring}, compositional visual reasoning (CVR)~\cite{zerroug2022benchmark}, synthetic visual reasoning test (SVRT)~\cite{fleuret2011comparing} with different attributes and rules (Fig.~\ref{fig:algo_framework}, Tab.~\ref{tab:selected_model}). Coupled with \textit{nsPE} reconfigurable arrays, \textit{BS} dataflow, and \textit{ST} mapping, \textit{adSCH} scheme ensures symbolic operations interleaved with neural operations with high throughput, enabling various kinds of VSA-based neurosymbolic workloads to be executed on \NAME with high efficiency and utilization, and adapt to different neuro-symbolic workload ratios and unpredictably changing workloads.

\section{Evaluation Results}
\label{sec:eval}
This section first introduces the detailed settings for evaluating our proposed \NAME framework (Sec.~\ref{subsec:exp_setup}), and then benchmarks our proposed \NAME algorithm optimization (Sec.~\ref{subsec:eval_algo}) and accelerator (Sec.~\ref{subsec:eval_hw}), demonstrating the practical of efficient and scalable neurosymbolic systems.

\subsection{Experimental Setup}
\label{subsec:exp_setup}
\textbf{Datasets.} 
To evaluate the achieved cognitive reasoning capability of CogSys, we conduct experiments on the commonly-used spatial-temporal reasoning RAVEN~\cite{raven}, I-RAVEN~\cite{iraven}, PGM~\cite{barrett2018measuring}, CVR~\cite{zerroug2022benchmark}, and SVRT~\cite{fleuret2011comparing}. The task performance is measured by the probabilistic abduction accuracy.

\textbf{Algorithm setup.} We evaluate \NAME on three state-of-the-art VSA-based neurosymbolic workloads, i.e., NVSA~\cite{hersche2023neuro}, MIMONet~\cite{menet2024mimonets}, and LVRF~\cite{hersche2024probabilistic}. 
Following~\cite{hersche2023neuro,menet2024mimonets,hersche2024probabilistic}, we determine the training hyperparameters based on the end-to-end reasoning performance on the validation set. 

\textbf{Baselines.} We consider several hardware baselines, including TX2, Xavier NX, RTX GPU, Xeon CPU, and ML accelerators (TPU, MTIA, Gemmini). Tab.~\ref{tab:eval_hw_baseline} lists their configurations.

\begin{figure}
\centering
\vspace{-5pt}
\includegraphics[width=\columnwidth]
{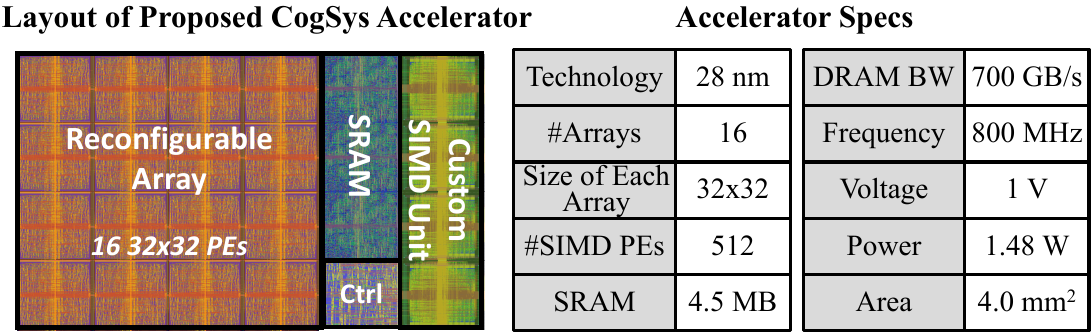}
        \vspace{-15pt}
            \caption{\textbf{\NAME accelerator.} \textnormal{The layout and performance specifications of our proposed \NAME accelerator.}}
        \label{fig:7.1chip}
        \vspace{-5pt}
\end{figure}

\textbf{Hardware setup.}
To evaluate energy and area of \NAME accelerator, we implement \NAME in RTL, synthesize using Synopsys Design Compiler~\cite{compiler} with 0.8~GHz, and place and route using Cadence Innovus~\cite{innovus} based on TSMC 28nm node. 
Fig.~\ref{fig:7.1chip} illustrates the layout and specifications of \NAME accelerator.
In addition, we develop a cycle-accurate simulator to estimate \NAME accelerator performance on different reasoning tasks.
The proposed \NAME accelerator consumes an area of 4.0~mm$^2$ and an average power consumption of 1.48~W. 
\emph{Compared with conventional systolic arrays that only support neural operations, \NAME provides reconfigurable support for neural and symbolic operations with only 4.8\% area overhead.}


\subsection{\NAME Algorithm Optimization Performance} 
\label{subsec:eval_algo}

\begin{table}[t!]
\centering
\caption{\textbf{Factorization accuracy comparison.} \textnormal{Factorization accuracy for object constituent attribute estimation across 14 scenarios.}}
\renewcommand*{\arraystretch}{1.3}
\setlength\tabcolsep{2.1pt}
\resizebox{\linewidth}{!}{%
\begin{tabular}{c|c|c|c|c|c|c|c|c}
\hline
\textbf{Test}    & \textbf{2$\times$2 Grid} & \textbf{3$\times$3 Grid}    & \textbf{Left-Right} & \textbf{Up-Down} & \textbf{Center} & \textbf{O-IC}       & \textbf{DistFour} & \textbf{Average} \\ \hline
\cite{langenegger2023memory} & 95.8\%            & 94.7\%               & 96.1\%              & 95.6\%           & 94.9\%          & 95.3\%              & 94.5\%                   & 95.3\%           \\ \hline
CogSys                & 95.7\%            & 95.2\%               & 96.1\%              & 95.7\%          & 95.3\%          & 95.5\%              & 94.4\%                   & 95.4\%           \\ \hline
\textbf{Test}       & \textbf{Constant} & \textbf{Progression} & \textbf{XOR}        & \textbf{AND}     & \textbf{OR}     & \textbf{Arithmetic} & \textbf{Distribution}    & \textbf{Average} \\ \hline
\cite{langenegger2023memory} & 93.3\%            & 93.5\%               & 93.9\%              & 93.7\%           & 93.5\%          & 93.1\%              & 92.7\%                   & 93.4\%          \\ \hline
CogSys                & 93.3\%            & 93.6\%               & 93.9\%              & 93.6\%           & 93.7\%          & 93.4\%              & 92.7\%                   & 93.5\%           \\ \hline
\end{tabular}
}
\label{tab:factorization_acc}
\end{table}

\textbf{Factorization accuracy.}
To assess the effectiveness of our factorization and stochasticity methods, we compare \NAME with the state-of-the-art factorizer~\cite{langenegger2023memory} across 14 test cases (Tab.~\ref{tab:factorization_acc}). The results show a slight improvement in factorization accuracy for object constituent attribute extraction.

\textbf{Reasoning accuracy.}
To evaluate \NAME algorithm optimization (Sec.~\ref{sec:algo_opt}), we benchmark it on five reasoning tasks in terms of the achieved accuracy (Sec.~\ref{subsec:exp_setup}). 
Tab.~\ref{tab:eval_acc} uses NVSA as an example and benchmarks on RAVEN, I-RAVEN, and PGM datasets, we observe that \NAME achieves comparable reasoning accuracy through factorization and injected stochasticity. Through quantization, \NAME enables 4.75$\times$ memory footprint savings as well as 7.71$\times$ area and 4.02$\times$ power savings (Tab.~\ref{tab:eval_quant}) under TSMC 28nm technology node. We get consistent observations in MIMONet and LVRF workloads under CVR and SVRT datasets as well.

\begin{table}[t!]
\centering
\caption{\textbf{\NAME algorithm optimization performance.} \textnormal{Compared with NVSA, \NAME exhibits comparable reasoning capability with smaller memory footprint requirement, achieved through the proposed factorization, stochasticity, and quantization techniques.}}
\renewcommand*{\arraystretch}{1.2}
\setlength\tabcolsep{2.1pt}
\resizebox{\linewidth}{!}{%
\begin{tabular}{c|ccc}
\hline
\multirow{1}{*}{\begin{tabular}[c]{@{}c@{}}Datasets\end{tabular}} 
                         & \multicolumn{1}{c|}{\begin{tabular}[c]{@{}c@{}}NVSA~\cite{hersche2023neuro}\end{tabular}} & \multicolumn{1}{c|}{\begin{tabular}[c]{@{}c@{}}\textbf{\NAME(+Factorization \& Stoch.)}\end{tabular}} & \begin{tabular}[c]{@{}c@{}}\textbf{\NAME (+Quant.)}\end{tabular}                        \\ \hline
RAVEN~\cite{raven}                    & \multicolumn{1}{c|}{98.5\%} & \multicolumn{1}{c|}{$\text{98.7}_{\pm\text{0.3}}$\%}                                                            & $\text{98.6}_{\pm\text{0.4}}$\%                                                                         \\ \hline
I-RAVEN~\cite{iraven}                  & \multicolumn{1}{c|}{99.0\%}   & \multicolumn{1}{c|}{$\text{99.0}_{\pm\text{0.3}}$\%}                                                            & $\text{98.8}_{\pm\text{0.4}}$\%                                                                   \\ \hline
PGM~\cite{barrett2018measuring}                  & \multicolumn{1}{c|}{68.3\%}   & \multicolumn{1}{c|}{$\text{68.6}_{\pm\text{0.8}}$\%}                                                            & $\text{68.4}_{\pm\text{1.0}}$\%                                                                    \\ \hline
\#Parameters                  & \multicolumn{1}{c|}{38~MB}   & \multicolumn{1}{c|}{32~MB}                                                            & 8~MB                                                             \\ \hline
\end{tabular}
}
\label{tab:eval_acc}
\end{table}

\begin{table}[t!]
\scriptsize
\centering
\caption{\textbf{Efficiency improvement from optimized precision.} \textnormal{\NAME optimizes NVSA algorithm to INT8 to enable hardware area and power savings while maintaining the reasoning capability.
}}
\renewcommand*{\arraystretch}{1.15}
\resizebox{\linewidth}{!}{%
    \begin{tabular}{cl|c|c|c}
    \hline
    \multicolumn{2}{c|}{Arithmetic Precision} & \multicolumn{1}{c|}{\textbf{FP32}} & \multicolumn{1}{c|}{\textbf{FP8}} & \multicolumn{1}{c}{\textbf{INT8}} \\ \hline
    \multicolumn{2}{c|}{\NAME Accuracy (NVSA=98.5\%)} & 98.9\% & 98.9\% & 98.7\% \\ \hline
    \multicolumn{1}{c|}{\multirow{2}{*}{\begin{tabular}[c]{@{}c@{}}Reconfigurable Array\\ 16 32$\times$32 PEs\end{tabular}}} & Area (mm$^2$) & 28.9 & 9.9 & 3.8 \\ \cline{2-5} 
    \multicolumn{1}{c|}{} & Power (mW) & 4468.5 & 1237.8 & 1104.6 \\ \hline
    \multicolumn{1}{c|}{\multirow{2}{*}{\begin{tabular}[c]{@{}c@{}}Custom SIMD Unit\\ 512 PEs\end{tabular}}} & Area (mm$^2$) & 2.01 & 0.28 & 0.21 \\ \cline{2-5} 
    \multicolumn{1}{c|}{} & Power (mW) & 297.0 & 64.8 & 80.4 \\ \hline
    \multicolumn{2}{c|}{Reconfig. Array Area Overhead vs. SA} & $<$1\% & 4.8\% & 12.1\% \\ \hline
\end{tabular}
}
\label{tab:eval_quant}
\vspace{-15pt}
\end{table}

\subsection{\NAME Accelerator Performance}
\label{subsec:eval_hw}

\begin{figure}[t!]
\centering
\includegraphics[width=\columnwidth]{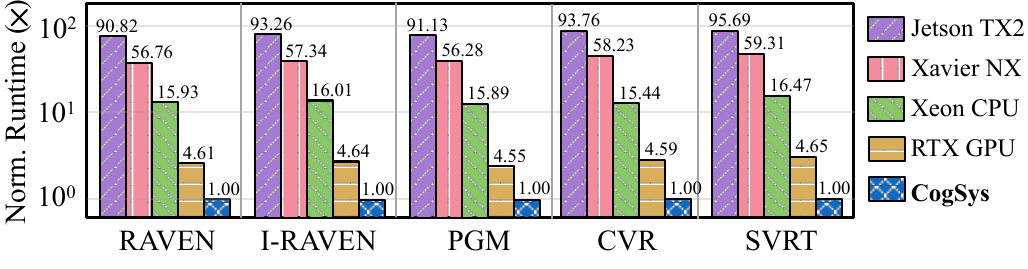}
\vspace{-0.2in}
        \caption{\textbf{End-to-end runtime improvement.} \textnormal{\NAME consistently outperforms Xeon CPU, RTX GPU, and edge SoCs (TX2, NX) in end-to-end runtime evaluated on five spatial-temporal reasoning tasks.}}
        \label{fig:eval_e2e_runtime}
        \vspace{-10pt}
\end{figure}

\textbf{Performance improvement.} We benchmark \NAME accelerator with RTX GPU, Xeon CPU, and edge SoCs (Jetson TX2, Xavier NX) for accelerating neurosymbolic algorithms on five reasoning tasks (Fig.~\ref{fig:eval_e2e_runtime}) with different difficulty levels.
For GPU baseline, for neuro kernels, we use Pytorch package that leverages CUDA and cuBLAS/cuDNN libraries; for symbolic kernels, we implement custom kernels optimized for vector-symbolic operations. The workload is tiled by CuDNN in Pytorch based on block sizes that fit well in GPU memory.
We observe that \NAME exhibits consistent speedup across datasets, e.g., 90.82$\times$/56.76$\times$ speedup over TX2 and NX, indicating its high efficiency and scalability capability. Furthermore, \NAME achieves real-time performance ($<$0.3~s)~\cite{hersche2023neuro} for solving logical reasoning tasks, indicating that \NAME is the \textit{first} to enable real-time neurosymbolic system with superior reasoning and generalization capability, offering a promising solution for future cognitive applications.

\begin{figure}[t!]
\centering
\includegraphics[width=\columnwidth]{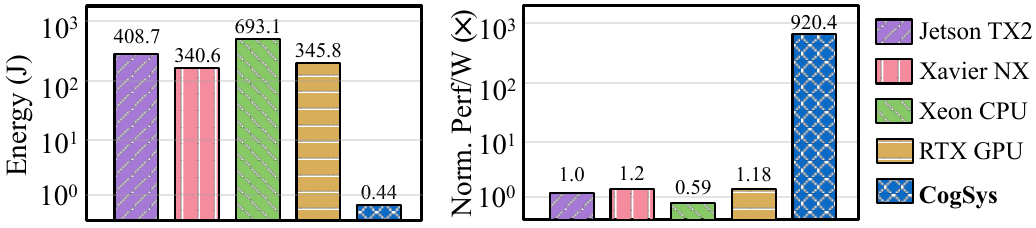}
\vspace{-0.2in}
        \caption{\textbf{Energy efficiency improvement.} \textnormal{\NAME consistently reduces energy consumption and improves performance per watt compared to CPU and GPUs, evaluated from five reasoning tasks.}}
        \label{fig:eval_energy}
        \vspace{-15pt}
\end{figure}

\textbf{Energy efficiency improvement.}
We benchmark \NAME accelerator on energy consumption and efficiency on five reasoning tasks (Fig.~\ref{fig:eval_energy}). We can observe that \NAME accelerator achieves two orders of energy efficiency than RTX GPU, Xeon CPU, TX2, and NX, indicating its efficiency and applicability to resource-constrained neurosymbolic systems. To further assess \NAME energy efficiency in long-term deployment, we conduct consecutive tests on CogSys using mixed workloads, incorporating both high-demand and low-activity periods, with 10-second idle intervals between scenarios. On average, CogSys achieves 730$\times$ energy efficiency compared to RTX GPU. Additionally, when compared to V100 and A100 GPUs, \NAME shows 4.43$\times$ and 1.43$\times$ speedup, with 748$\times$ and 241$\times$ energy efficiency, respectively.



\begin{figure}[t!]
\centering
\vspace{-5pt}
\includegraphics[width=\columnwidth]{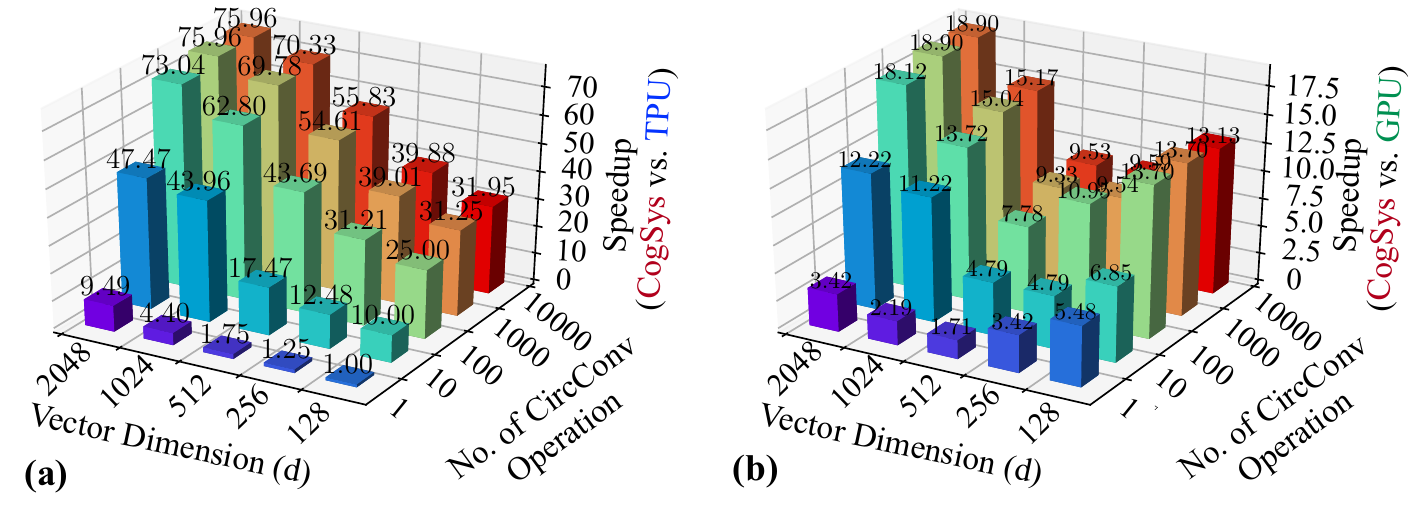}
        \vspace{-0.2in}
        \caption{\textbf{Improved efficiency over TPU/GPU.}
        \textnormal{Speedup comparison of circular convolution on CogSys, TPU-like systolic array and GPU, \NAME shows up to 75.96$\times$ and 18.90$\times$ runtime improvement.}}
        \label{fig:tpu_gpu}
        \vspace{-5pt}
\end{figure}

\textbf{Comparison with TPU/GPU.} 
We benchmark symbolic circular convolution over TPU-like SA (with the same number of PEs) and GPU under different vector dimensions and number of operations (Fig.~\ref{fig:tpu_gpu}). 
We observe \NAME  reconfigurable array achieves up to 75.96$\times$ and 18.90$\times$ speedup over TPU-like SA and GPU, and is effective in both low-dimension and high-dimension vector-symbolic operations.

\begin{figure}[t!]
\centering
\includegraphics[width=\columnwidth]{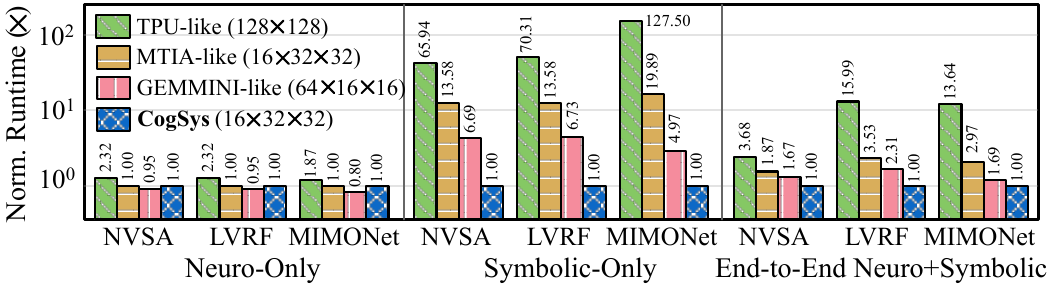}
        \vspace{-0.15in}
        \caption{\textbf{Improved efficiency over ML accelerators.}
        \textnormal{Speedup comparison of neural, symbolic, and end-to-end neurosymbolic system over TPU~\cite{jouppi2021ten}, Gemmini~\cite{genc2021gemmini}, and MTIA~\cite{firoozshahian2023mtia}-like architecture.}}
        \label{fig:ml_acc_comp}
        \vspace{-5pt}
\end{figure}

\textbf{Comparison with ML accelerators.} We benchmark the runtime of neural and symbolic operations on TPU~\cite{jouppi2021ten}, Gemmini~\cite{genc2021gemmini}, and MTIA~\cite{firoozshahian2023mtia}-like architecture over different neurosymbolic models and tasks (Fig.~\ref{fig:ml_acc_comp}). For a fair comparison, we keep all hardware configurations with the same number of PEs. Compared with current ML accelerators, we observe that \NAME achieves similar performance in neural operations, while exhibiting superior symbolic operation efficiency thus end-to-end speedup in neurosymbolic systems. Additionally, we compare \NAME with hyperdimensional computing accelerator~\cite{ibrahim2024efficient} across models and tasks and observe a 7.2$\times$ average speedup. This improvement is mainly due to the lack of efficient neuro and symbolic support and circular convolution handling in hyperdimensional computing architectures.

\begin{figure}[t!]
\centering
\vspace{-5pt}
\includegraphics[width=\columnwidth]{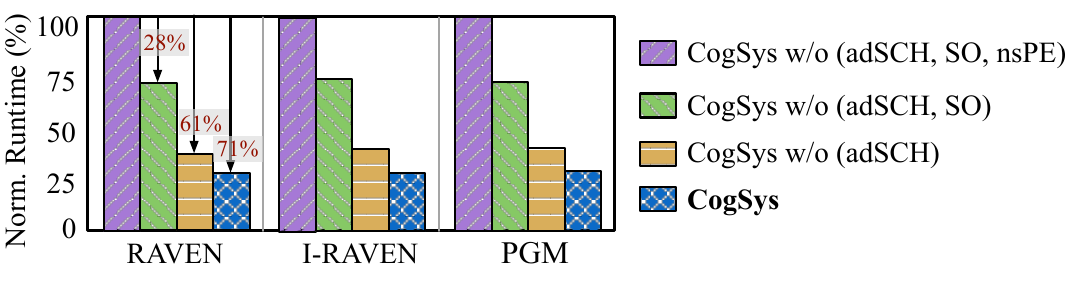}
\vspace{-0.15in}
        \caption{\textbf{Ablation Study on \NAME Accelerator Techniques}. \textnormal{The runtime achieved by \NAME w/o the adaptive scheduling (adSCH), scalable array (SO), and reconfigurable PE (nsPE) across tasks.}}
        \label{fig:eval_ablation}
        \vspace{-10pt}
\end{figure}

\textbf{Ablation study on the proposed hardware techniques.} As illustrated in Sec.~\ref{sec:accelerator} and Sec.~\ref{sec:sched}, \NAME features reconfigurable neuro/symbolic PE with bubble streaming dataflow and spatial-temporal mapping, scalable array architecture, and adaptive scheduling strategy to reduce compute latency and memory footprint for neural and symbolic kernels. To verify the effectiveness of our proposed methods, we summarize the runtime of \NAME w/o the scheduling, scalable architecture, and reconfigurable PE in Fig~\ref{fig:eval_ablation}. In particular, the proposed scheduling strategy can trim down the runtime by 28\% on average. Additionally, with the proposed scalable array and reconfigurable PE, the runtime reduction ratio can be further enlarged to 61\% and 71\%, indicating that both proposed techniques are necessary for our \NAME accelerator to achieve the desired efficient and scalable reasoning capability.

\textbf{Ablation study of necessity of co-design.} To the best of our knowledge, our proposed CogSys, as an algorithm-hardware co-design framework, is the first that has achieved efficient and scalable on-device neurosymbolic-based system. To verify the necessity of such co-design strategy, we summarize the runtime of our \NAME w/o the proposed algorithm optimization or hardware techniques in Tab.~\ref{tab:ablation_codesign}. Specifically, with our proposed \NAME algorithm optimization, we can trim down the runtime to 89.5\% as compared to NVSA~\cite{hersche2023neuro} on the same Xavier NX  hardware and RAVEN task. Moreover, with both proposed \NAME algorithm optimization and accelerator, the runtime can be reduced to 1.76\%, indicating the necessity of the co-design strategy of \NAME framework.

\begin{table}[t!]
\centering
\caption{\textbf{Ablation study of necessity of co-design.} \textnormal{The normalized runtime achieved by \NAME framework w/o the proposed algorithm optimization or hardware techniques on different tasks.}}
\renewcommand*{\arraystretch}{1.2}
\setlength\tabcolsep{1.6pt}
\resizebox{\linewidth}{!}{%
\begin{tabular}{c|ccccc}
\hline
Neurosymbolic Cognitive Solution & \multicolumn{5}{c}{Normalized Runtime (\%) on} \\
Algorithm @ Hardware & RAVEN~\cite{raven}  & I-RAVEN~\cite{iraven}     & PGM~\cite{barrett2018measuring} &  CVR~\cite{zerroug2022benchmark} & SVRT~\cite{fleuret2011comparing}    \\ \hline
NVSA~\cite{hersche2023neuro} @ Xavier NX &  100     &   100    & 100  & 100 & 100    \\ \hline
\textbf{\NAME Algorithm @ Xavier NX} &  89.5\%     &  88.9\%     &  90.7\%  & 87.6\%	& 88.4\%   \\ \hline
\textbf{\NAME Algorithm @ \NAME Accelerator}  &   1.76\%    & 1.74\%     &   1.78\% & 1.72\% & 1.69\%   \\ \hline
\end{tabular}
}
\label{tab:ablation_codesign}
\vspace{-15pt}
\end{table}

\section{Related Work}
\label{sec:related}
 
\textbf{Neurosymbolic AI.}
Neurosymbolic AI holds significant potential for enhancing trustworthiness, reasoning, and robustness of next-generation cognitive applications where agents can make decisions in an explainable manner, and intelligence is pervasively embedded in human-AI interactions~\cite{zhang2020alphazero,badreddine2022logic,hohenecker2020ontology,dongneural,hersche2023neuro,pryor2022neupsl,manhaeve2021neural,yang2020neurasp,yi2020clevrer}. 
Current neurosymbolic research mostly focuses on algorithms; however, the lack of attention to its inefficiency on off-the-shelf hardware may hinder neurosymbolic AI development in the long run. \emph{\NAME thus takes the first step to understand neurosymbolic architectural and system characteristics and proposes a co-design framework to make it more efficient and deployable at scale.}

\textbf{Accelerators for emerging applications.}
With the slowdown of technology scaling, custom architecture is a pragmatic approach to ensure simultaneous improvements in performance and efficiency. Beyond DNNs~\cite{genesys,zhou2022ml,tambe202216,tpuv4,shao2019simba,brainwave,ramachandran2024algorithm,tambe2020algorithm}, hardware acceleration has been found effective in emerging applications such as genome sequencing \cite{fujiki2020seedex, fujiki2018genax}, graph~\cite{gao2023mega,shah2022dpu}, mobile vision~\cite{liu2020systolic,liu2022s2ta,wan2024h3dfact}, drone~\cite{krishnan2022automatic,chang202373,krishnan2022roofline}, robotics~\cite{neuman2023roboshape,mayoral2024robotperf,liu2022energy,hao2024orianna,liu2021robotic}, privacy and security~\cite{f1,samardzic2022craterlake, geelen2022basalisc,nabeel2023cofhee}, etc.
\textit{Despite the presence of these accelerators, \NAME is the first proposal to offer reconfigurable support for both neural and symbolic kernels, facilitating efficient and scalable neurosymbolic systems.}
%

\vspace{-2pt}
\section{Conclusion}
\label{sec:conclusion}

To enable efficient and scalable neurosymbolic AI towards real-time cognitive applications, we propose CogSys, the first algorithm-hardware co-design framework dedicated to accelerating neurosymbolic AI. \NAME identifies the unique opportunities for neurosymbolic acceleration, including efficient factorization, reconfigurable neural/symbolic PE, bubble streaming dataflow, and adaptive scheduler, leveraging which we develop algorithm optimizations and dedicated accelerator. We believe \NAME can open up an exciting perspective toward efficient and scalable cognitive reasoning systems at scale.
\section*{Acknowledgements}
This work was supported in part by CoCoSys, one of seven centers in JUMP 2.0, a Semiconductor Research Corporation (SRC) program sponsored by DARPA.


\bibliographystyle{IEEEtranS}
\bibliography{refs}

\end{document}